\documentclass[%
preprint,
 amsmath,amssymb,
 aps,
]{revtex4-2}

\usepackage{graphicx}
\usepackage{dcolumn}
\usepackage{bm}

\usepackage{latexsym}
\usepackage[dvipsnames]{xcolor}
\usepackage{graphicx}
\usepackage{newtxtext}
\usepackage{newtxmath}
\usepackage{natbib}
\usepackage{subcaption}
\usepackage{hyperref}
\hypersetup{
    colorlinks = true,
    urlcolor   = blue,
    citecolor  = black,
}

\newcommand{\RomanNumeralCaps}[1]
\linenumbers

\begin{document}

\title{Optimal transient growth and transition to turbulence in the MHD pipe flow subject to a transverse magnetic field}

\author{Yelyzaveta Velizhanina}
\affiliation{
  Université Libre de Bruxelles, Faculté des Sciences, Physique des Systèmes Dynamiques, CP231, 1050 Brussels, Belgium
}%
\author{Bernard Knaepen}
\affiliation{
  Université Libre de Bruxelles, Faculté des Sciences, Physique des Systèmes Dynamiques, CP231, 1050 Brussels, Belgium
}%

\date{\today}%

\begin{abstract}
We consider the influence of a transverse magnetic field on the transient growth of perturbations in a liquid-metal circular pipe flow with an electrically insulating or conducting wall. In this configuration, the mean flow profile and the amplification of perturbations are strongly affected by the applied magnetic field, leading to a rich dynamical landscape depending on its intensity. The analysis is performed for Reynolds numbers 5000 and 10 000, close to the transitional regime for moderate values of the Hartmann number, a non-dimensional parameter proportional to the applied magnetic field's intensity. Aside from a slight modification of hydrodynamic optimal perturbations at very small Hartmann numbers, we observe three other characteristic topologies of optimal perturbations depending on the intensity of the magnetic field. Their growth mechanisms differ, with the lift-up effect dominating at low Hartmann numbers and the Orr-mechanism becoming increasingly important as the magnetic field intensity is increased. In particular, we show in the intermediate regime of Hartmann numbers how transient growth occurs in two stages, with initial growth through the lift-up effect followed by a further increase of energy through the Orr-mechanism. We also conduct three-dimensional nonlinear simulations to track the time evolution of optimal perturbations, illustrating their nonlinear growth and eventual breakdown to a sustained turbulent state or the return of the system to a laminar state.
\end{abstract}

\maketitle

\section{Introduction}\label{sec:introduction}
The behavior of electrically conducting fluids under the influence of magnetic fields is crucial in numerous technological and industrial applications, such as steel manufacturing and processing, semi-conductor crystal growth, as well as in the development of heat transfer systems or plasma facing components for future nuclear fusion reactors.

Applied magnetic fields can profoundly alter the dynamics of an electrically conducting flow. This includes Joule damping, the creation of thin boundary layers and velocity jets in the mean velocity profile, and the promotion of two-dimensional motion by the suppression of three-dimensional variations along the magnetic field lines \cite{Hunt1965,SommeriaMoreau1982,KnaepenMoreau2008,Davidson2001}. Stability-wise, magnetic fields can therefore suppress transition by damping perturbations, yet they can also introduce new types of instabilities through modifications in velocity profiles \cite{Buhler2017, BlishchikKenjeres2022}.

In this work, we address the laminar-turbulent transition in the magnetohydrodynamic (MHD) pipe flow subject to a transverse magnetic field with perfectly insulating and perfectly conducting electromagnetic boundary conditions. In constrast to its hydrodynamic counterpart, this flow is linearly unstable to three-dimensional perturbations \cite{VelizhaninaKnaepen2023}.
Notably, the instability occurs solely in the presence of an electrically conducting wall and a sufficiently strong magnetic field, for Reynolds numbers exceeding the global critical value of 45 230. However, classical experiments performed by Hartmann and Lazarus \cite{HartmannLazarus1937}, supported by results of direct numerical simulations (DNS) \cite{SatakeKunugiSmolentsev2002,KrasnovThessBoeckZhaoZikanov2013}, indicate that depending on the intensity of the magnetic field, the MHD pipe flow exhibits full or patterned turbulence at significantly lower values of $Re$. This discrepancy suggests that the observed laminar-turbulent transition is subcritical and is likely caused by the breakdown of initially small flow disturbances as they achieve finite amplitude through algebraic transient growth.

The transient growth of perturbations in MHD flows has been the subject of several studies addressing mostly the channel flow \cite{AiriauCastets2004,KrasnovRossiZikanovBoeck2008,DongKrasnovBoeck2012,DongKrasnovBoeck2015,DongKrasnovBoeck2016}, the duct flow \cite{KrasnovZikanovRossiBoeck2010,HuLiuZhangNi2015,CassellsVoPotheratSheard2019,CamobrecoPotheratSheard2023} or boundary layer flows \cite{CamobrecoPotheratSheard2020,BourcyVelizhaninaPavlenkoKnaepen2022}.
Particularly relevant to the present work are studies dedicated to the transient growth of disturbances in a rectangular duct with perfectly insulating walls in the presence of transverse magnetic field. The first comprehensive analysis, conducted in Ref.~\cite{KrasnovZikanovRossiBoeck2010}, focused on moderate values of the Hartmann number. In this flow regime, the optimal perturbation initially takes the form of oblique vortices, localized in the Shercliff boundary layers. At the time of maximum amplification, it is transformed into oblique streaks. Based on the topology of optimal disturbances, the authors have identified the lift-up effect as the driver of perturbation growth. This central mechanism in the transition to turbulence in many shear flows causes the perturbation to evolve into longitudinal streaks by lifting low-velocity fluid away from the wall and pushing high-velocity fluid toward the wall \cite{Landahl1980}.
The transient growth analysis of the MHD duct flow has later been extended to the range of high Hartmann numbers in Ref.~\cite{CassellsVoPotheratSheard2019}. In the presence of sufficiently strong magnetic field, the optimal perturbation emerged in the form of quasi-two-dimensional spanwise vortices. In this regime, the amplification of the optimal perturbation is achieved through the Orr-mechanism, which consists in the transport of momentum down the mean momentum gradient \cite{Orr1907}. Following this finding, the authors suggested that quasi-two-dimensional turbulence in the presence of a sufficiently strong magnetic field \cite{SommeriaMoreau1982,PotheratKlein2014}, can be initiated directly from a quasi-two dimensional laminar state (rather than from an intermediate three-dimensional state). This path to turbulence has not been tested by the means of DNS, although a quasi-two-dimensional transition scenario has been identified in Ref.~\cite{CamobrecoPotheratSheard2023} for the MHD duct flow with lateral walls.

To date, no detailed study of the transient growth of perturbations in the MHD pipe flow with transverse magnetic field has been performed. In (hydrodynamic) pipe Poiseuille flow, the maximum amplification occurs for a two-dimensional disturbance in the form of purely streamwise vortices \cite{Bergstrom1993,SchmidHenningson1994}. Similar to the duct flow, these evolve into longitudinal streaks through the lift-up effect that can be destabilized by a small three-dimensional disturbance, as demonstrated in the DNS performed in Ref.~\cite{Zikanov1996}. The same behavior has later been observed in other hydrodynamic shear flows, such as plane Poiseuille flow \cite{ReddySchmidBaggettHenningson1998} and square duct flow \cite{BiauSoueidBottaro2008}. In this context, the influence of a {\em streamwise} magnetic field on the transient growth in a liquid-metal flow in a perfectly conducting pipe has been considered in Ref.~\cite{Akerstedt1995}.
In this case, the magnetic field does not modify the base velocity profile, nor does it affect the evolution of streamwise streaks. Therefore, analogous to the hydrodynamic case, the maximum linear growth is achieved for a two-dimensional disturbance amplified via the lift-up effect. However, in the nonlinear regime, the three-dimensional secondary instability of streamwise streaks exhibits magnetic damping. This property has been investigated in Ref.~\cite{DongKrasnovBoeck2012} in the context of the MHD channel flow with perfectly conducting walls and a streamwise magnetic field, showing that optimal perturbations were fully damped at sufficiently high Hartmann numbers.

The MHD pipe flow with transverse magnetic field considered in this article has a much richer dynamical landscape compared to the streamwise magnetic field configuration considered in Ref.~\cite{Akerstedt1995}. The main reason is of course the significant departure of the base flow from Poiseuille profile as the intensity of the magnetic field is increased. Key characteristics of the MHD base flow profile include the formation of two new types of boundary layers: the Hartmann and Roberts layers, which develop near the wall where it is perpendicular or parallel to the applied magnetic field, respectively. From the point of view of transient growth, a transverse magnetic field also damps streamwise vortices and directly affects the dominant growth mechanism present in the hydrodynamic pipe flow. As we shall see below, this leads to the emergence of multiple flow regimes with different optimal perturbation topologies and a gradual transition from the lift-up effect to the Orr-mechanism as the primary means of perturbation amplification.

Our work is organized as follows. The physical model and problem formulation are introduced in Section~\ref{sec:formulation}. The description and validation of the numerical method used for the transient growth analysis are then given in Section~\ref{sec:numerical-method}. Section~\ref{sec:results} presents the results of this study and is divided in three parts. In the first one, we present the global characteristics of transient growth as we vary the intensity of the applied magnetic field. In the second part, we describe in detail the topology of the optimal perturbations identified in the different regimes encountered and highlight the corresponding amplification mechanisms.
The third part of Section~\ref{sec:results} then focuses on nonlinear simulations that illustrate the breakdown to turbulence of the previously described optimal perturbations. Section~\ref{sec:conclusions} closes this article and contains our main conclusions.

\section{Formulation}\label{sec:formulation}
\begin{figure}
  \centering
  \includegraphics[width=0.75\textwidth,trim={7.5cm 3.8cm 3cm 8cm},clip]{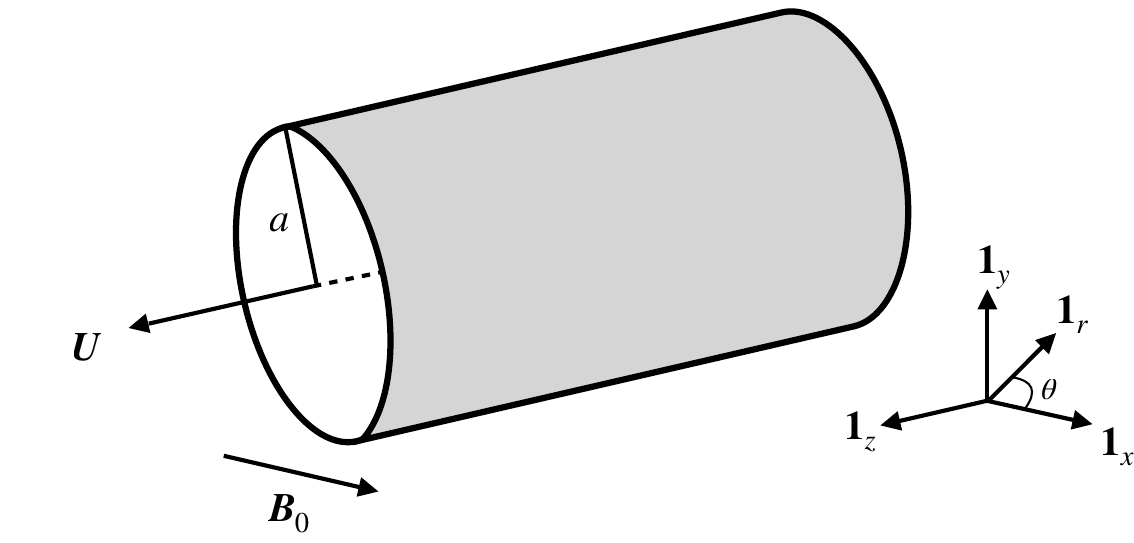}
  \caption{Schematic representation of the MHD pipe flow subject to a transverse magnetic field.}
  \label{figure:geometry}
\end{figure}

We consider a viscous, incompressible, electrically conducting fluid with density $\rho$, kinematic viscosity $\nu$, and electric conductivity $\sigma$, contained in a pipe with circular cross-section and wall conductivity $\sigma_w$. The system is subject to a constant pressure gradient and uniform transverse magnetic field $\boldsymbol{B}_0= B_0 \boldsymbol{1}_x$. The sketch of flow's geometry is shown in Fig.~\ref{figure:geometry}.

In this work, we model the flow dynamics using the quasi-static approximation, which provides an accurate description of most laboratory liquid-metal flows. It assumes that the induced magnetic field adjusts to variations in the velocity nearly instantaneously and is negligible compared to the applied magnetic field. As characteristic scales of the problem, we choose the radius of the pipe $a$, the maximum mean flow velocity $U$, and the magnitude of the applied magnetic field $B_0 = \lvert \boldsymbol{B}_0 \rvert$. The non-dimensional variables are therefore defined through:
\begin{equation}
\setlength{\arraycolsep}{10pt}
\renewcommand{\arraystretch}{1.6}
\begin{array}{ccc}
  \boldsymbol{u} \to U\boldsymbol{u}, & p \to \rho U^2 p, & \phi\to a U B_0\phi,\\
  \boldsymbol{x}\to a\boldsymbol{x}, & t\to aU^{-1}t, & \boldsymbol{B}_0\to B_0\boldsymbol{1}_x, \\
\end{array}
\end{equation}
where $\boldsymbol{u}$, $p$ and $\phi$ denote, respectively, the fluid velocity, pressure, electric potential, and $\boldsymbol{1}_x$ is the unit vector aligned with $\boldsymbol{B}_0$.

The quasi-static MHD equations are then given by (see Ref.~\cite{MuellerBuhler2001} for details):
\begin{gather}
  \nabla\cdot\boldsymbol{u} = 0,\label{eq:mhd-equations-1}\\
  \frac{\partial\boldsymbol{u}}{\partial t} + \left( \boldsymbol{u}\cdot\nabla \right)\boldsymbol{u} = -\nabla p + \frac{1}{Re}\nabla^2 \boldsymbol{u} + \frac{Ha^2}{Re} \left( -\nabla \phi + \boldsymbol{u}\times \boldsymbol{1}_x \right)\times \boldsymbol{1}_x, \label{eq:mhd-equations-2}\\
  \nabla^2 \phi = \nabla\cdot\left( \boldsymbol{u}\times \boldsymbol{1}_x \right).\label{eq:mhd-equations-3}
\end{gather}
Here, the non-dimensional parameters $Re = U a / \nu$ and $Ha = a B_0 \sqrt{\sigma / \rho \nu}$ are respectively the Reynolds number and the Hartmann number. At the wall, $\boldsymbol{u}$ satisfies the no-slip condition,
\begin{equation}
  \boldsymbol{u} = 0 \quad \text{at} \quad r = 1.
\end{equation}
In this study, we use two types of electromagnetic boundary conditions, obtained by considering the behavior of the electric current $\boldsymbol{j} =  -\nabla \phi + \boldsymbol{u}\times \boldsymbol{1}_x$ at the wall.
When the wall is perfectly insulating, i.e. $\sigma_w = 0$, no current can leave the flow domain, and the electric potential $\phi$ is subject to a Neumann boundary condition,
\begin{equation}
  \frac{\partial \phi}{\partial r} = 0 \quad \text{at} \quad r = 1.
\end{equation}
When the wall is perfectly conducting, i.e. $\sigma_w = \infty$, the electric potential must be constant at the wall to avoid infinite electric currents. We therefore have:
\begin{equation}
  \phi = C \quad \text{at} \quad r = 1,
\end{equation}
where $C$ is an arbitrary constant.

To linearize Eqs.~\eqref{eq:mhd-equations-1}-\eqref{eq:mhd-equations-3}, we decompose the instantaneous flow variables $\boldsymbol{u}$, $p$, and $\phi$ into a steady two-dimensional base state and a three-dimensional infinitesimal perturbation. In cylindrical coordinates $(r,\theta,z)$, this can be expressed as follows:
\begin{equation}\label{eq:decomposed-flow-fields}
  \renewcommand{\arraystretch}{1.6}
  \begin{array}{c}
    \boldsymbol{u}\left(r, \theta, z, t\right) = U\left( r,\theta \right)\boldsymbol{1}_z + \boldsymbol{u}'\left(r, \theta, t\right)e^{i\alpha z}, \\ p\left(r, \theta, z, t\right) = P\left(r, \theta, z \right) + p'\left(r, \theta, t\right)e^{i\alpha z}, \\ \phi\left(r, \theta, z, t\right) = \Phi\left(r, \theta \right) + \phi'\left(r, \theta, t\right)e^{i\alpha z},
  \end{array}
  \end{equation}
where $\alpha$ denotes the axial wavenumber of periodicity along the axis of the pipe.

The equations governing the base state of the flow are given by:
\begin{gather}
  (\nabla^2 - Ha^2) U + Ha^2 \left( \sin{\theta}\frac{\partial}{\partial r} + \frac{\cos{\theta}}{r}\frac{\partial}{\partial\theta} \right) M = Re K,\label{eq:base-equation-1}\\
  \nabla^2 \Phi - \left( \sin{\theta}\frac{\partial}{\partial r} + \frac{\cos{\theta}}{r}\frac{\partial}{\partial\theta} \right) U = 0\label{eq:base-equation-2},
\end{gather}
where $K = \partial P/\partial z$ is the streamwise pressure drop. Upon substituting Eqs.~\eqref{eq:decomposed-flow-fields} into Eqs.~\eqref{eq:mhd-equations-1}-\eqref{eq:mhd-equations-3}, we obtain, after linearization, the following perturbation equations:
\begin{gather}
  \nabla\cdot\boldsymbol{u}' = 0,\label{eq:pert-equation-1}\\
  \frac{\partial\boldsymbol{u}'}{\partial t} + \left( \boldsymbol{U}\cdot\nabla \right)\boldsymbol{u}' + \left( \boldsymbol{u}'\cdot\nabla \right)\boldsymbol{U} = -\nabla p' + \frac{1}{Re}\nabla^2\boldsymbol{u}' + \frac{Ha^2}{Re}\left( -\nabla\phi' + \boldsymbol{u}'\times\boldsymbol{1}_x \right)\times\boldsymbol{1}_x,\label{eq:pert-equation-2}\\
  \nabla^2 \phi' = \nabla\cdot\left( \boldsymbol{u}'\times \boldsymbol{1}_x \right),\label{eq:pert-equation-3}
\end{gather}
where $\nabla = \partial / \partial r \boldsymbol{1}_r + r^{-1}\partial / \partial \theta \boldsymbol{1}_\theta + i\alpha\boldsymbol{1}_z$.

The main purpose of transient growth analysis is to determine the initial conditions of the linear problem that experience the largest growth according to a given norm. To that end, we define the inner product and the associated norm, based on the kinetic energy of disturbance, through:
\begin{equation}\label{eq:def-of-inner-product}
   \langle \boldsymbol{u}', \boldsymbol{u}' \rangle = \lVert \boldsymbol{u}' \rVert^2 = E_{\boldsymbol{u}'}\left( t \right) = \frac{L_\alpha}{8}\int_0^1\int_{0}^{2\pi} \left( u'^*_r u'_r + u'^*_\theta u'_\theta + u'^*_z u'_z \right)r d\theta dr,
\end{equation}
where $L_\alpha = 2\pi/\alpha$ is perturbation wavelength and $*$ denotes complex conjugate. Consequently, the maximum amplification of kinetic energy at time $t$, attained for some optimal initial condition $\boldsymbol{u}'_0$ is defined by:
\begin{equation}
  G\left( Re, Ha, \sigma_w, \alpha, t \right) = \max_{\boldsymbol{u}'_0}\frac{E_{\boldsymbol{u}'}\left( t \right)}{E_{\boldsymbol{u}'}\left( 0 \right)}.\label{eq:def-of-maximum-growth}
\end{equation}
For given values of $Re$, $Ha$ and $\sigma_w$, the amplification $G$ depends on the particular choice of streamwise wavenumber $\alpha$ and the time at which the kinetic energy is measured.
The optimal amplification $G_\alpha\left( Re, Ha, \sigma_w, t \right)$ is then defined by maximizing $G$ over all values of $\alpha$ at a given time $t$. Finally, the global optimal amplification $G_{\textrm{opt}}\left( Re, Ha, \sigma_w \right)$ is computed by maximizing $G$ over all possible values of $\alpha$ and $t$. $G_{\textrm{opt}}$ represents the largest kinetic energy amplification that can ever be achieved by a perturbation of the form defined in Eq.~\eqref{eq:decomposed-flow-fields}, and the corresponding perturbation is referred to as the global optimal perturbation.

To perform the transient growth analysis, we follow the optimization procedure described in Ref.~\cite{SchmidHenningson1994}. All the details of our numerical implementation of the method are discussed in the next section, while the code used for the nonlinear analysis of the evolution of optimal perturbations is described in Section~\ref{sec:nonlinear}.
\section{Numerical method}\label{sec:numerical-method}
The base-flow Eqs.~\eqref{eq:base-equation-1} and \eqref{eq:base-equation-2} and the perturbation Eqs.~\eqref{eq:pert-equation-1}-\eqref{eq:pert-equation-3} are discretized using a Chebyshev-Fourier collocation method. For that purpose, we expand the flow variables over suitable cardinal functions and seek solutions in physical space $\boldsymbol{x} = \left( r,\theta \right)$. To reduce the clustering of the grid points near the origin of the pipe and eliminate the coordinate singularity at $r=0$, we exploit the so-called ``rotate-and-reflect'' symmetry \cite{Trefethen2000}. This is done by initially discretizing all flow quantities in a redundant virtual domain $\left[ -1, 1 \right]\times [ -\pi, \pi )$ and subsequently imposing the following conditions:
\begin{equation}\label{eq:rotate-and-reflect}
  \renewcommand{\arraystretch}{1.3}
  \begin{aligned}
    p'\left( r,\theta,t \right) & = p'\left( -r,\theta\pm\pi,t \right),\\
    u_r'\left( r,\theta,t \right) & = -u_r'\left( -r,\theta\pm\pi,t \right),\\
    u_\theta'\left( r,\theta,t \right) & = -u_\theta'\left( -r,\theta\pm\pi,t \right),\\
    u_z'\left( r,\theta,t \right) & = u_z'\left( -r,\theta\pm\pi,t \right),\\
    \phi'\left( r,\theta,t \right) & = \phi'\left( -r,\theta\pm\pi,t \right).
  \end{aligned}
  \end{equation}
Additionally, we take advantage of the symmetric nature of the base flow and split the original problem into two independent sets of linear equations for perturbations that are either symmetric or anti-symmetric with respect to $r=0$. The new computational domain becomes $\left[ 0, 1 \right]\times [ 0, \pi )$, and the overall storage requirement is reduced by a factor of $4$.

The computational grid contains $N = N_r N_\theta$ grid points, where $N_r$ and $N_\theta$ denote the number of Chebyshev and Fourier collocation points, respectively. Evaluating continuous quantities $\boldsymbol{u}'$, $p'$ and $\phi'$ at the grid points yields the vectors
\begin{equation}\label{eq:discrete-variables}
  \renewcommand{\arraystretch}{1.5}
  \begin{array}{c}
    \hat{\boldsymbol{u}} = \left( u'_{r,0},\dots, u'_{r,N-1}, u'_{\theta,0},\dots, u'_{\theta,N-1}, u'_{z,0},\dots, u'_{z,N-1} \right)^T,\\
    \hat{p} = \left( p'_{0},\dots, p'_{N-1} \right)^T,\\
    \hat{\phi} = \left( \phi'_{0},\dots, \phi'_{N-1} \right)^T.
  \end{array}
  \end{equation}
Consequently, the discretized Eqs.~\eqref{eq:pert-equation-1}-\eqref{eq:mhd-equations-3} can be expressed in matrix-vector notations as follows:
\begin{gather}
  F\hat{\boldsymbol{u}} = 0,\label{eq:full-disc-1}\\
  \frac{d\hat{\boldsymbol{u}}}{dt} = S\hat{\boldsymbol{u}} + M\hat{p} + Q\hat{\phi},\label{eq:full-disc-2}\\
  D\hat{\phi} = E\hat{\boldsymbol{u}},\label{eq:full-disc-3}
\end{gather}
where $S$ and $D$ are square matrices of respective sizes $3N$ and $N$, $M$ and $Q$ are rectangular matrices of size $3N\times N$, and $F$ and $E$ are rectangular matrices of size $N\times 3N$. The boundary conditions are integrated into the system as linear equations.

By solving Eq.~\eqref{eq:full-disc-3} subject to the suitable electromagnetic boundary condition for the electric potential, we eliminate $\hat{\phi}$ and close the problem in terms of hydrodynamic variables only. The resulting system contains $4N$ unknowns, including the boundary values, and is given by:
\begin{gather}
  F\hat{\boldsymbol{u}} = 0,\label{eq:disc-1}\\
  \frac{d\hat{\boldsymbol{u}}}{dt} = L\hat{\boldsymbol{u}} + M\hat{p},\label{eq:disc-2}
\end{gather}
with $L = S + Q\left( D^{-1} E \right)$. To eliminate the pressure and further reduce the computational requirements, we use the method of algebraic reduction \cite{BoikoNechepurenko2008,BoikoNechepurenkoSadkane2012}. To that end, we compute the QR-decomposition of matrices $M$ and $F^*$ (here $*$ denotes conjugate transpose),
\setlength{\jot}{0.5cm}
\begin{gather}
  M = M_Q M_R = \begin{bmatrix}
    \smash{\underbrace{\vphantom{\rule[-0.4em]{0pt}{0pt}}M_{Q,1}}_{N}} & \smash{\underbrace{\vphantom{\rule[-0.4em]{0pt}{0pt}}M_{Q,2}}_{N}}
  \end{bmatrix}\begin{bmatrix}
    M_{R,1} \\ 0
  \end{bmatrix}
  \hspace{-0.1cm}
  \setlength\arraycolsep{0.02cm}
  \begin{matrix}
    \left.\vphantom{\begin{matrix}0\end{matrix}}\right\} & N \\ \left.\vphantom{\begin{matrix}0\end{matrix}}\right\} & 2N
  \end{matrix}\,,\\
  F^* = F_Q F_R = \begin{bmatrix}
    \smash{\underbrace{\vphantom{\rule[-0.4em]{0pt}{0pt}}F_{Q,1}}_{N}} & \smash{\underbrace{\vphantom{\rule[-0.4em]{0pt}{0pt}}F_{Q,2}}_{N}}
  \end{bmatrix}\begin{bmatrix}
    F_{R,1} \\ 0
  \end{bmatrix}
  \hspace{-0.1cm}
  \setlength\arraycolsep{0.02cm}
  \begin{matrix}
    \left.\vphantom{\begin{matrix}0\end{matrix}}\right\} & N \\ \left.\vphantom{\begin{matrix}0\end{matrix}}\right\} & 2N
  \end{matrix}\,,
\end{gather}
\vspace{0.25cm}

\noindent where $M_Q$ and $F_Q$ are unitary matrices with orthonormal columns, and $M_{R,1}$ and $F_{R,1}$ are nonsingular upper triangular matrices of size $N$. Consequently, the new set of $2N$ unknowns $\hat{\boldsymbol{v}}$ is defined by:
\begin{equation}
  \hat{\boldsymbol{v}} = F^*_{Q,2}\hat{\boldsymbol{u}}\quad\text{and}\quad\hat{\boldsymbol{u}} = F_{Q,2}\hat{\boldsymbol{v}}.
\end{equation}
Substituting these expressions into Eq.~\eqref{eq:disc-2} and multiplying the result by $M^*_{Q,2}$ leads to the following equation for $\hat{\boldsymbol{v}}$:
\begin{equation}\label{eq:reduced-ivp}
  \frac{d\hat{\boldsymbol{v}}}{dt} = H\hat{\boldsymbol{v}},
\end{equation}
where $H = A^{-1}L'$, $A = M^*_{Q,2}F_{Q,2}$ and $L' = M^*_{Q,2} L F_{Q,2}$. The solution to this equation is given by:
\begin{equation}
  \hat{\boldsymbol{v}}\left( t \right) = \exp{\left( H t \right)} \hat{\boldsymbol{v}}_0,
\end{equation}
where $\hat{\boldsymbol{v}}_0$ is an arbitrary initial condition. In terms of the original variables, we then have:
\begin{equation}
  \hat{\boldsymbol{u}}\left( t \right) = F_{Q,2} \exp{\left( H t \right)} F^*_{Q,2} \hat{\boldsymbol{u}}_0.
\end{equation}
Using Eq.~\eqref{eq:def-of-inner-product}, we write:
\begin{equation}
  \frac{E_{\hat{\boldsymbol{u}}}\left( t \right)}{E_{\hat{\boldsymbol{u}}}\left( 0 \right)} = \frac{\langle W F_{Q,2} \exp{\left( H t \right)} F^*_{Q,2} \hat{\boldsymbol{u}}_0, F_{Q,2} \exp{\left( H t \right)} F^*_{Q,2} \hat{\boldsymbol{u}}_0 \rangle}{\langle W \hat{\boldsymbol{u}}_0, \hat{\boldsymbol{u}}_0 \rangle},
\end{equation}
where the Hermitian positive definite matrix $W$ accounts for the non-uniformity of the computational grid and is defined by:
\begin{equation}
  W = \begin{pmatrix}
    1 & 0 & 0 \\
    0 & 1 & 0 \\
    0 & 0 & 1
  \end{pmatrix}
  \otimes \begin{pmatrix}
    w_0    & 0      & \dots & 0      \\
    0      & w_1    & \dots & 0      \\
    \vdots & \ddots &       & \vdots \\
    0      & 0      & \dots & w_{N-1}
  \end{pmatrix},
\end{equation}
with the coefficients $\left\{ w_0,\dots w_{N-1} \right\}$ denoting Chebyshev-Fourier quadrature weights. The maximum amplification of perturbation kinetic energy $G\left( Re, Ha, \sigma_w, \alpha, t \right)$, occurring for some optimal initial condition $\hat{\boldsymbol{u}}_0$, is then given by:
\begin{equation}
  \begin{split}
    G\left( Re, Ha, \sigma_w, \alpha, t \right) & = \max_{\hat{\boldsymbol{u}}_0}{\frac{\langle W F_{Q,2} \exp{\left( H t \right)} F^*_{Q,2} \hat{\boldsymbol{u}}_0, F_{Q,2} \exp{\left( H t \right)} F^*_{Q,2} \hat{\boldsymbol{u}}_0 \rangle}{\langle W \hat{\boldsymbol{u}}_0, \hat{\boldsymbol{u}}_0 \rangle}} \\ & = \max_{\hat{\boldsymbol{u}}_0}{\frac{\langle \exp{\left( \tilde{H} t \right)} \tilde{\boldsymbol{u}}_0, \exp{\left( \tilde{H} t \right)} \tilde{\boldsymbol{u}}_0 \rangle}{\langle \tilde{\boldsymbol{u}}_0, \tilde{\boldsymbol{u}}_0 \rangle}} = \lVert \exp{\left( \tilde{H} t \right)} \rVert^2_2 = \varsigma^2,
  \end{split}
\end{equation}
where $\tilde{\boldsymbol{u}}_0 = W^{1/2}\hat{\boldsymbol{u}}_0$, $\tilde{H} = W^{1/2} F_{Q,2} H F^*_{Q,2} W^{-1/2}$, and $\varsigma$ is the largest singular value of $\exp{\left( \tilde{H} t \right)}$ \cite{SchmidHenningson2001}. Using the eigenvalue decomposition of $H = P \Lambda P^{-1}$, where $\Lambda$ and $P$ are composed of the eigenvalues and eigenvectors of $H$, respectively, the operator exponential can be constructed as:
\begin{equation}
  \exp{\left( \tilde{H} t \right)} = \tilde{P} \exp{\left( \Lambda t \right)} \tilde{P}^{-1},
\end{equation}
where $\tilde{P} = W^{1/2} F_{Q,2} P$.

\begin{figure}
  \centering
  \begin{subfigure}[t]{0.475\textwidth}
    \centering
    \includegraphics[width=\textwidth,trim={0 0 0 0},clip]{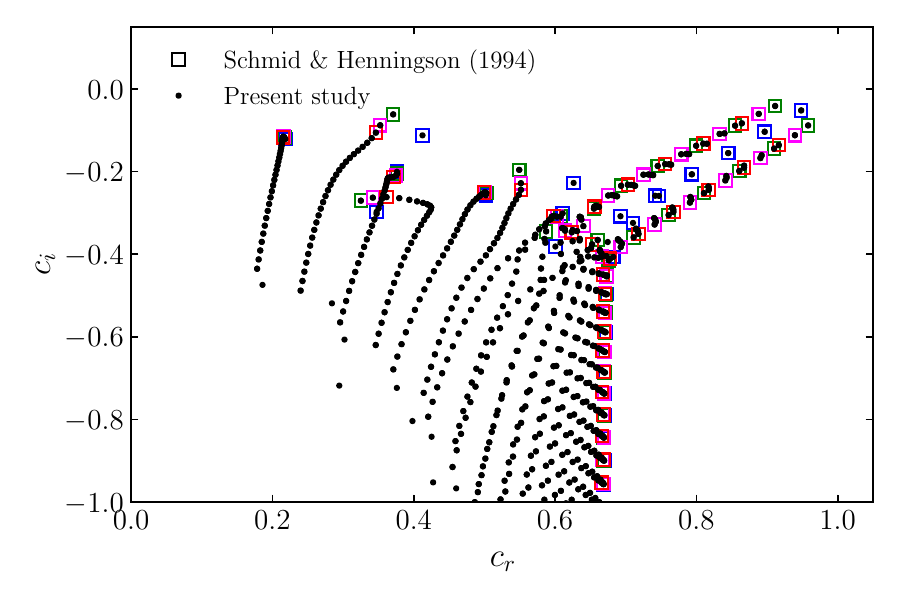}
  \end{subfigure}
  \begin{subfigure}[t]{0.4425\textwidth}
    \centering
    \includegraphics[width=\textwidth,trim={1cm 0 0 0},clip]{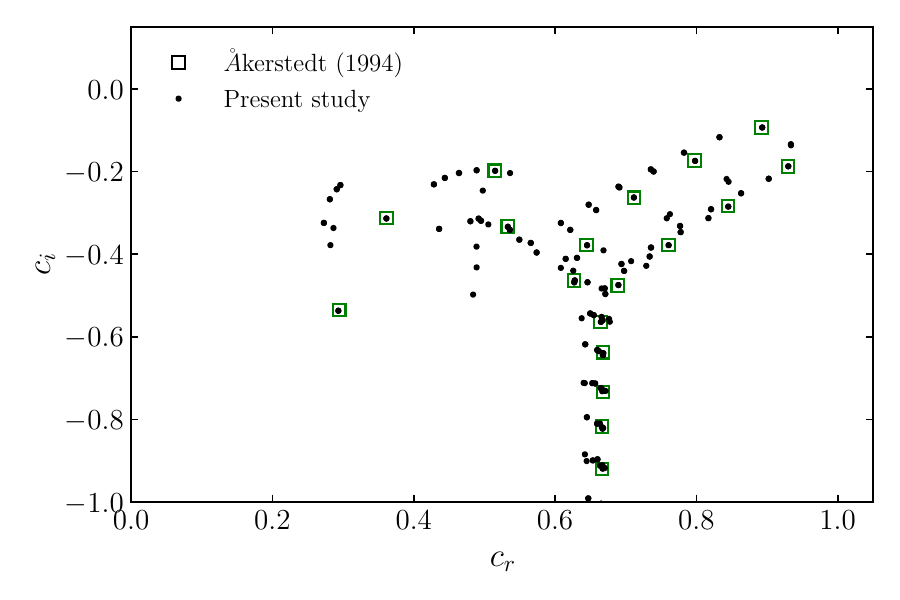}
  \end{subfigure}
  \caption{Eigenvalue spectra pipe Poiseuille flow (left) and MHD pipe flow with streamwise magnetic field (right). $N_r N_\theta = 30\times 20$ in both cases. Results obtained in this study are compared with those reported in Refs.~\cite{SchmidHenningson1994} and \citep{Akerstedt1995} for $n=0$ ({\color{blue}blue}), $n=1$ ({\color{Green}green}), $n=2$ ({\color{magenta}magenta}) and $n=3$ ({\color{red}red}).}
  \label{figure:valid_spectra}
\end{figure}

\begin{figure}
  \centering
  \includegraphics[width=0.65\textwidth,trim={0 0 0 0},clip]{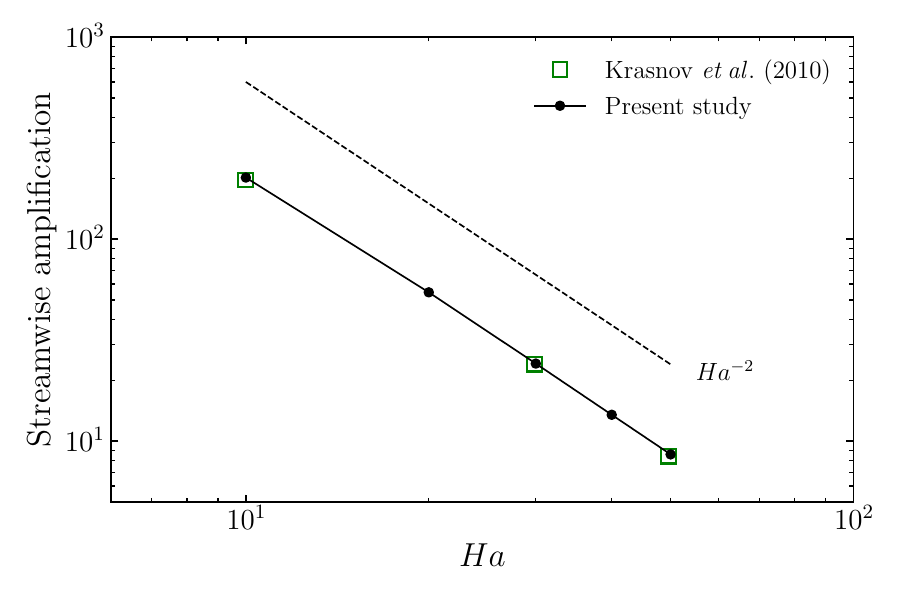}
  \caption{Global maximum amplification as a function of the Hartmann number for perturbations with $\alpha = 0$ in the MHD flow contained in a square perfectly insulating duct with transverse magnetic field. Results obtained in this study are compared with those reported in Ref.~\cite{KrasnovZikanovRossiBoeck2010}.}
  \label{figure:valid_krasnov_jfm_2010_5}
\end{figure}

The above procedure to compute the amplification $G$ has been implemented in an in-house solver relying on the NumPy library \cite{Harris2020}. To validate our solver, we have considered several test cases available in the literature. The construction of the linear operator $H$ appearing in Eq.~\eqref{eq:reduced-ivp} is validated using the results of the linear stability analysis of pipe Poiseuille flow \cite{SchmidHenningson1994}, as well as MHD pipe flow with streamwise magnetic field \cite{Akerstedt1995}.
The respective eigenvalue spectra are shown in Fig.~\ref{figure:valid_spectra}. As our code is 2D, it captures all the eigenvalues for different azimuthal wavenumbers at once (compared to a 1D code where the azimuthal wavenumber $n$ is a parameter). For the two problems considered and for all the values $n$ available in Refs.~\cite{SchmidHenningson1994} and \cite{Akerstedt1995}, we obtain an excellent match.
In a second stage of validation, we have compared our predictions for the maximum amplification $G$ in the hydromagnetic flow contained in a perfectly insulating square  duct with a transverse magnetic field with those available in Ref.~\cite{KrasnovZikanovRossiBoeck2010} and obtained excellent agreement. As an illustration, we show in Fig.~\ref{figure:valid_krasnov_jfm_2010_5} the maximum global amplification for pure streamwise perturbations ($\alpha=0$) obtained using our formulation with 35 Chebyshev collocation points in every direction.
\section{Results}\label{sec:results}
In this work we focus our attention on the transient growth of perturbations at $Re=5000$ and vary the Hartmann number from 0 to 100. This choice of Reynolds number is motivated by the fact that we are mostly interested in the transitional regime for moderate Hartmann numbers. The same Reynolds number was also used in this regime for the detailed analysis of patterned turbulence in Ref.~\cite{KrasnovThessBoeckZhaoZikanov2013}. We also provide some results for $Re=10 \,000$, but we did not observe principal differences in the existence of the different regimes of optimal modes as we explore the range $0 \le Ha \le 100$. Note that in all the cases considered, the flow is linearly stable from the point of view of modal analysis as it can only become unstable when considering conducting walls and for $Re>45230$ (see Ref.~\cite{VelizhaninaKnaepen2023}).
\subsection{Maximum amplification}
\begin{figure}
  \centering
  \includegraphics[width=0.9\textwidth]{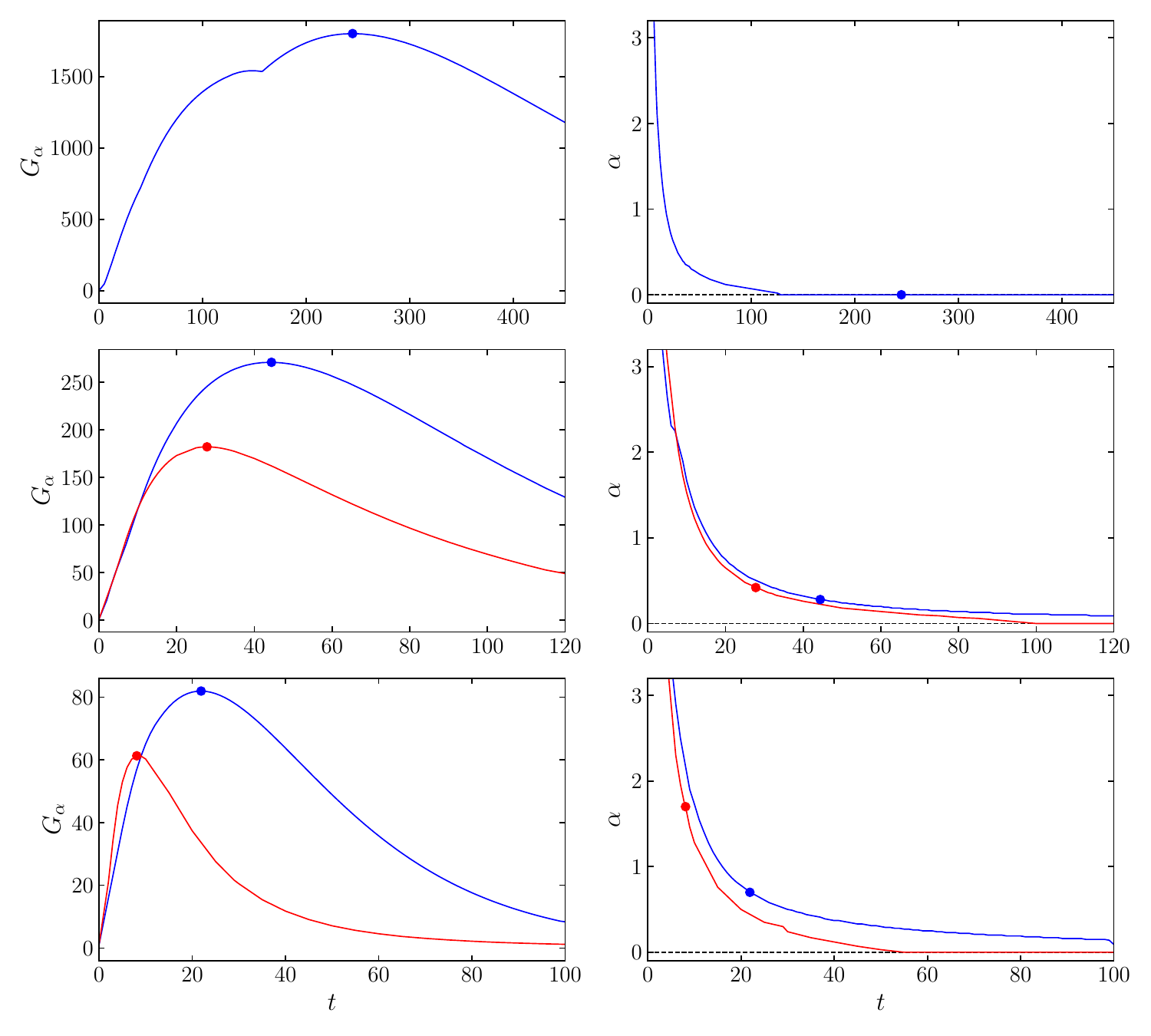}
  \caption{Optimal amplification $G_\alpha$ and corresponding optimal axial wavenumber $\alpha$ as functions of time $t$. The results are shown for $Re=5000$ and for $Ha=0$ (top), $H=10$ (middle) and $Ha=20$ (bottom). For $Ha > 0$, the {\color{blue}blue} and {\color{red}red} lines correspond, respectively, to $\sigma_w=0$ and $\sigma_w=\infty$. For each curve, the marker indicates the global maximum $G_{\textrm{opt}}\left( Re, Ha, \sigma_w \right)$ and the corresponding $\alpha_{\textrm{opt}}$.}
  \label{figure:G-alpha-vs-time}
\end{figure}

We first discuss the main trends of transient growth. In Fig.~\ref{figure:G-alpha-vs-time}, the optimal amplification $G_\alpha\left( Re, Ha, t \right)$ and the corresponding wavenumber $\alpha$ are depicted as functions of $t$ for $Re=5000$ and for different values of $Ha$ in the perfectly insulating and perfectly conducting cases.
The damping effect of the magnetic field is evident as the maximum amplification is already reduced by more than one order of magnitude for $Ha=20$ and very limited for $Ha>50$ (see Table~\ref{tab:optVals} for more data). As expected, maximum amplification is smaller in the perfectly conducting case due to the increased Joule damping resulting from stronger currents flowing through the pipe's wall. For all values of $Ha$, the optimal wavenumber $\alpha$ decays monotonically with $t$ and the initial growth rate of $G_\alpha$ increases with $Ha$.
\begin{table}[b]
  \begin{ruledtabular}
  \begin{tabular}{ccccc}
    $\sigma_w$ & $Ha$ & $G_{\textrm{opt}}$ & $\alpha_{\textrm{opt}}$ & $t_{\textrm{opt}}$ \\
  \colrule
  0        & 0    & 1801.7 & 0    & 244.8 \\
           & 1    & 1685.3 & 0    & 239.3 \\
           & 5    & 657.4  & 0.03 & 107.4 \\
           & 10   & 271.1  & 0.28 & 44.4  \\
           & 20   & 82     & 0.7  & 21.9  \\
           & 50   & 10.8   & 2.13 & 8.1   \\
           & 75   & 5.5    & 2.4  & 8.    \\
           & 100  & 4.6    & 2.6  & 8.6   \\
  $\infty$ & 1    & 1664.3 & 0    & 237.4 \\
           & 5    & 564.8  & 0.06 & 93.9  \\
           & 10   & 182.3  & 0.42 & 27.8  \\
           & 20   & 61.3   & 1.7  & 8.1   \\
           & 50   & 9.5    & 3.57 & 2.6   \\
           & 75   & 3.2    & 3.43 & 1.4   \\
           & 100  & 1.3    & 4    & 1     \\
  \end{tabular}
  \end{ruledtabular}
  \caption{Global optimal amplification for $Re=5000$ and different values of the Hartmann number $Ha$ in the perfectly insulating and perfectly conducting cases. The values of optimal time $t_{\textrm{opt}}$ and optimal axial wavenumber $\alpha_{\textrm{opt}}$ are determined with an accuracy of 0.1 and 0.01, respectively.}
  \label{tab:optVals}
  \end{table}

\begin{figure}
    \centering
    \includegraphics[width=0.75\textwidth]{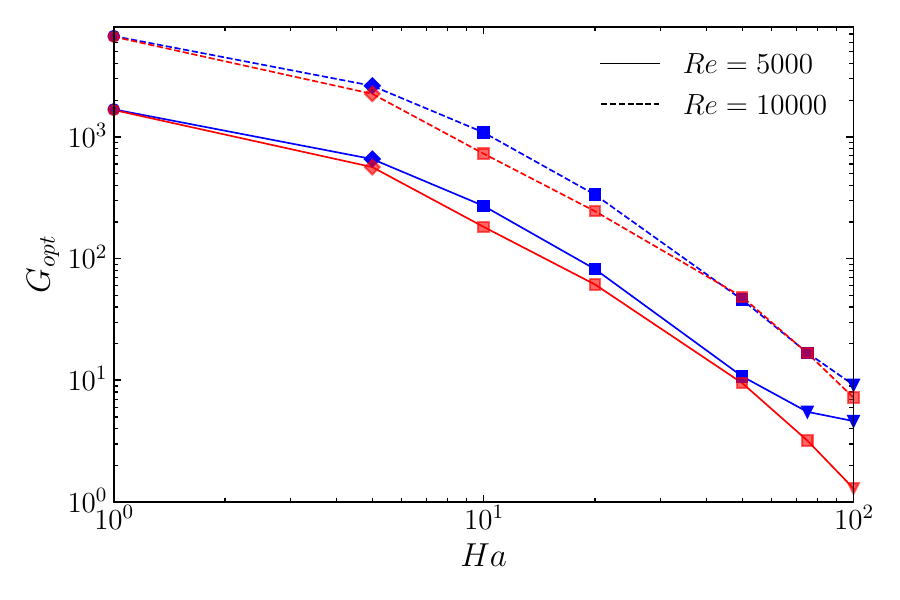}
    \caption{Optimal amplification $G_{\textrm{opt}}$ as a function of Hartmann number in the perfectly insulating ({\color{blue}blue}) and perfectly conducting ({\color{red}red}) cases. The solid and dashed curves correspond to $Re=5000$ and $Re=10\,000$, respectively. The different markers represent the different topologies ($\circ$ - hydrodynamic, $\Diamond$ - low-$Ha$ regime, $\square$ - intermediate-$Ha$ regime, $\bigtriangledown$ - high-$Ha$ regime) of optimal perturbations described in section~\ref{sec:topology}. }
    \label{figure:optimal-growth}
\end{figure}

Fig.~\ref{figure:optimal-growth} displays the global optimal amplification $G_{\textrm{opt}}$ as a function of the Hartmann number for both wall conductivities.  For moderate values of the Hartmann number ($10 \le Ha \le 50$), we observe a scaling of $G_{\textrm{opt}} \propto Ha^{-2}$ (this contrasts with the MHD duct flow with transverse magnetic field where the scaling is $G_{\textrm{opt}} \propto Ha^{-1.5}$ \citep{KrasnovZikanovRossiBoeck2010}). For higher Hartmann numbers, the departure of $G_{\textrm{opt}}$ from this scaling law can be attributed to the emergence of a different regime of optimal modes (see Section~\ref{sec:topology} for details).

\begin{figure}
  \centering
  \includegraphics[width=\textwidth]{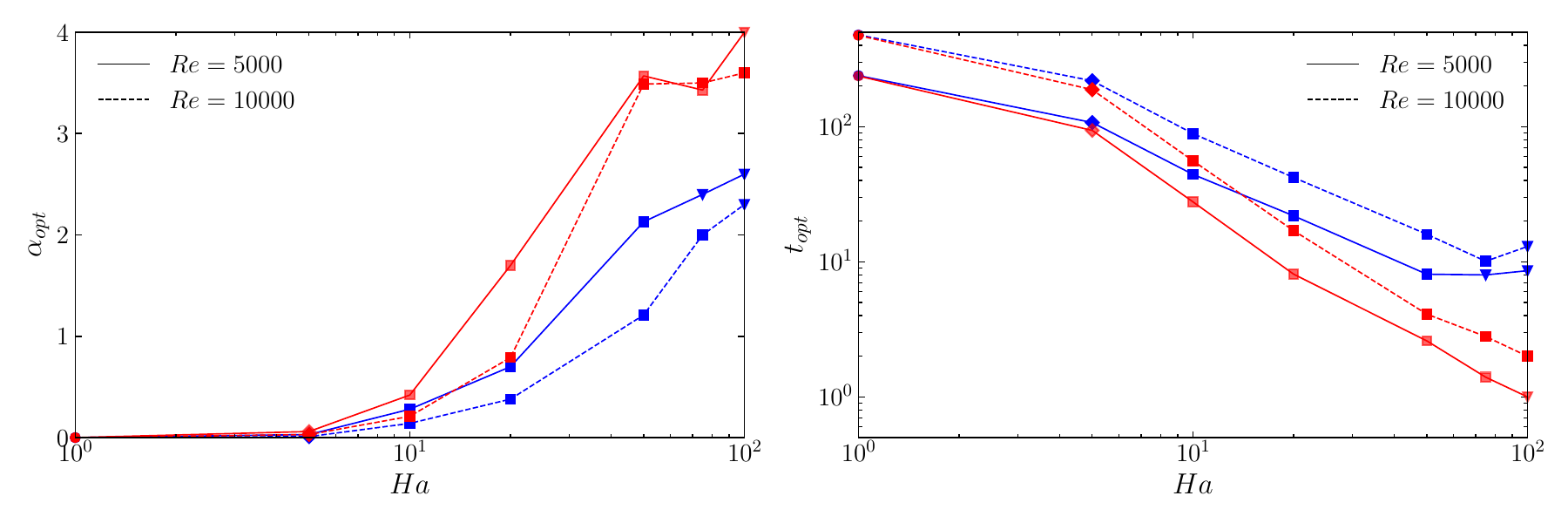}
  \caption{Optimal axial wavenumber $\alpha_{\textrm{opt}}$ (left) and global optimal time $t_{\textrm{opt}}$ (right) as functions of the Hartmann number in the perfectly insulating ({\color{blue}blue}) and perfectly conducting ({\color{red}red}) cases. The solid and dashed curves correspond to $Re=5000$ and $Re=10\,000$, respectively. The different markers represent the different topologies of optimal perturbations (see Fig.~\ref{figure:optimal-growth} for details).}
  \label{figure:optimal-params}
\end{figure}

The optimal wavenumbers $\alpha_{\textrm{opt}}$ and optimal times $t_{\textrm{opt}}$ for the different Hartmann numbers considered are shown in Fig.~\ref{figure:optimal-params}. Compared to the hydrodynamic pipe flow where global optimal perturbations are streamwise independent, the MHD pipe flow with transverse magnetic field is characterized by optimal perturbations with non-zero $\alpha$. These perturbations also reach their maximum amplification earlier in time when the Hartmann number is increased.
In terms of $\alpha_{\textrm{opt}}$ and $t_{\textrm{opt}}$, the difference between the perfectly insulating and perfectly conducting cases becomes significant for $Ha > 5$.
For $\sigma_w = \infty$, the optimal disturbance is then characterized by shorter axial wavelength, attaining maximum amplification at earlier times than for $\sigma_w = 0$.
\subsection{Topology of optimal disturbances and mechanisms of transient growth}\label{sec:topology}
In this section, we investigate the structure of the global optimal perturbations and the respective mechanisms leading to their transient growth. We find three types of global optimal modes, distinct from those in the hydrodynamic case, occurring for ``low", ``moderate" and ``high'' values of $Ha$. They are observed within similar ranges of Hartmann numbers in the perfectly insulating and perfectly conducting cases.

In pipe Poiseuille flow, the global optimal perturbation consists in a pair of streamwise counter-rotating vortices near the axis of the pipe, with nearly 100\% of total kinetic energy contained in the perpendicular velocity components $u_r$ and $u_\theta$ \citep{Bergstrom1993,SchmidHenningson1994}. Over time, the perturbation evolves into pairs of high and low-velocity longitudinal structures. They are referred to as streamwise streaks to indicate the dominance of $u_z$ over the other velocity components. For that perturbation, all the velocity components are antisymmetric with respect to the centerline of the pipe.

\subsubsection{Low values of \textit{Ha}}
For weak magnetic fields, the MHD global optimal perturbation is very similar to the hydrodynamic one, except for a modulation with large wavelength in the streamwise direction. This behavior is also observed in the MHD duct flow for $Ha\leq 1$ \citep{CassellsVoPotheratSheard2019} and is not furhter investigated here.

\begin{figure}
  \centering
  \begin{subfigure}[t]{0.4\textwidth}
      \centering
      \includegraphics[width=\textwidth,trim={24cm 2cm 15cm 1cm},clip]{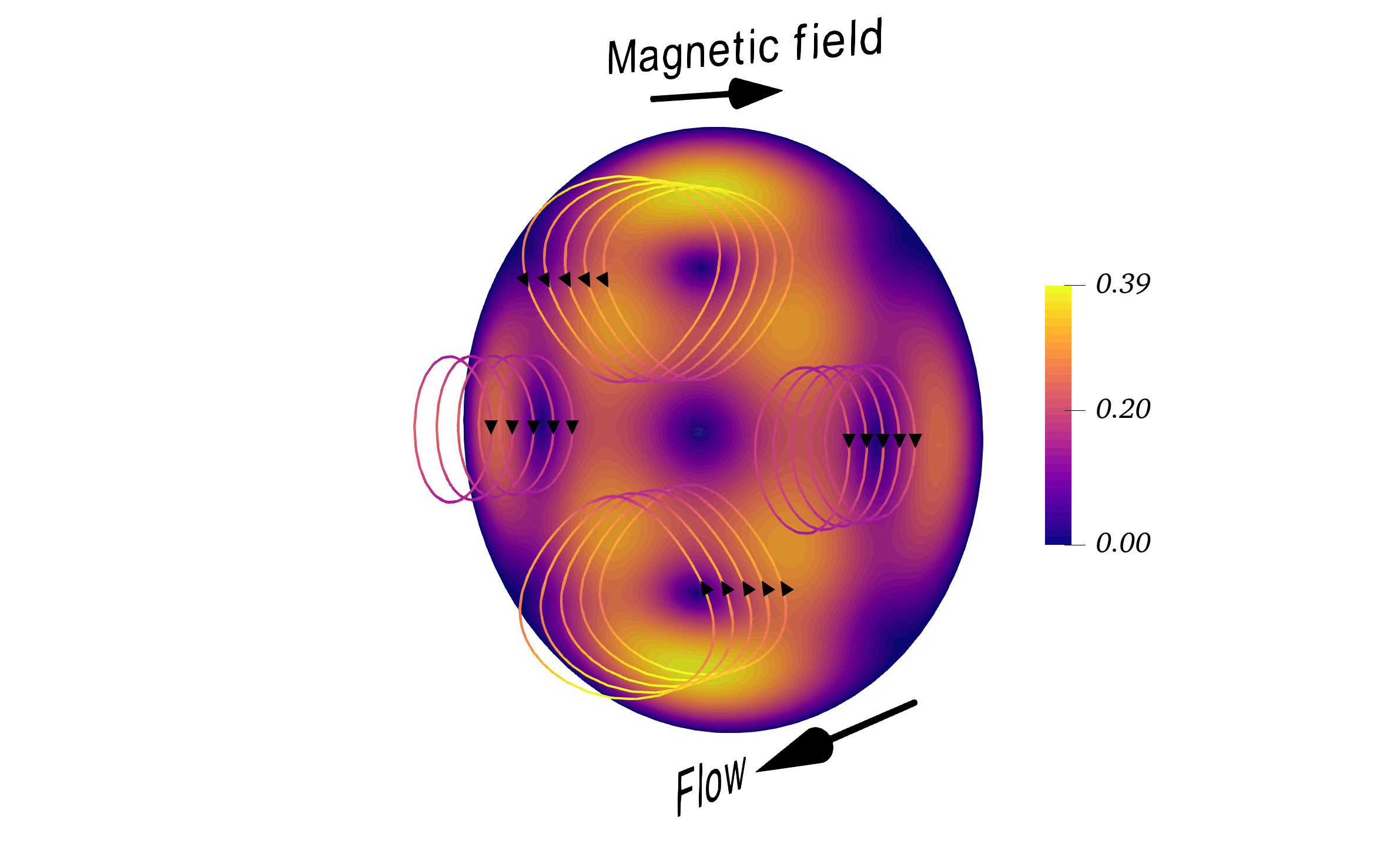}
  \end{subfigure}
  \begin{subfigure}[t]{0.4\textwidth}
      \centering
      \includegraphics[width=\textwidth,trim={20cm 2cm 20cm 1cm},clip]{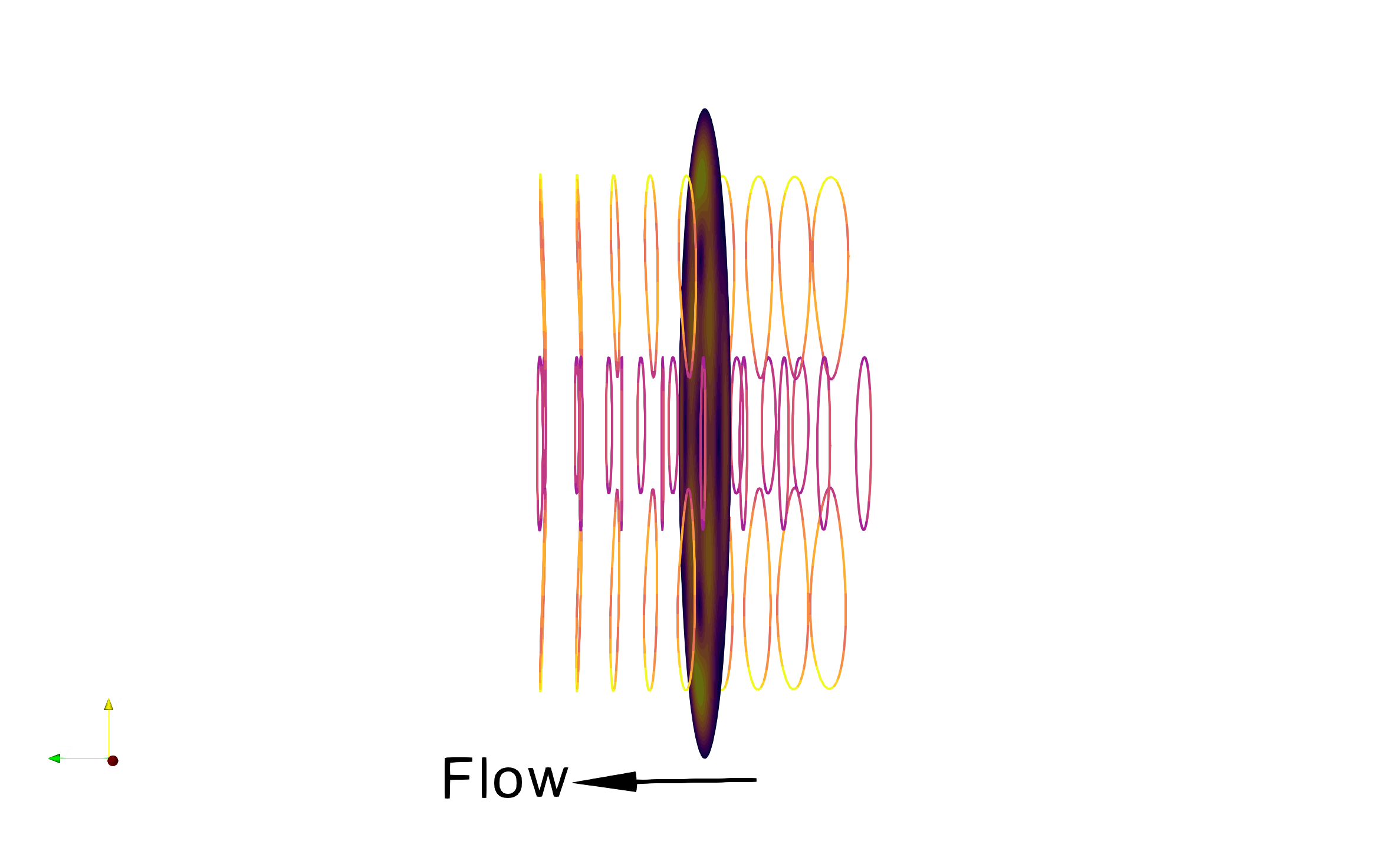}
  \end{subfigure}
  \vskip\baselineskip
  \begin{subfigure}[t]{0.4\textwidth}
    \centering
    \includegraphics[width=\textwidth,trim={25cm 4cm 15cm 3cm},clip]{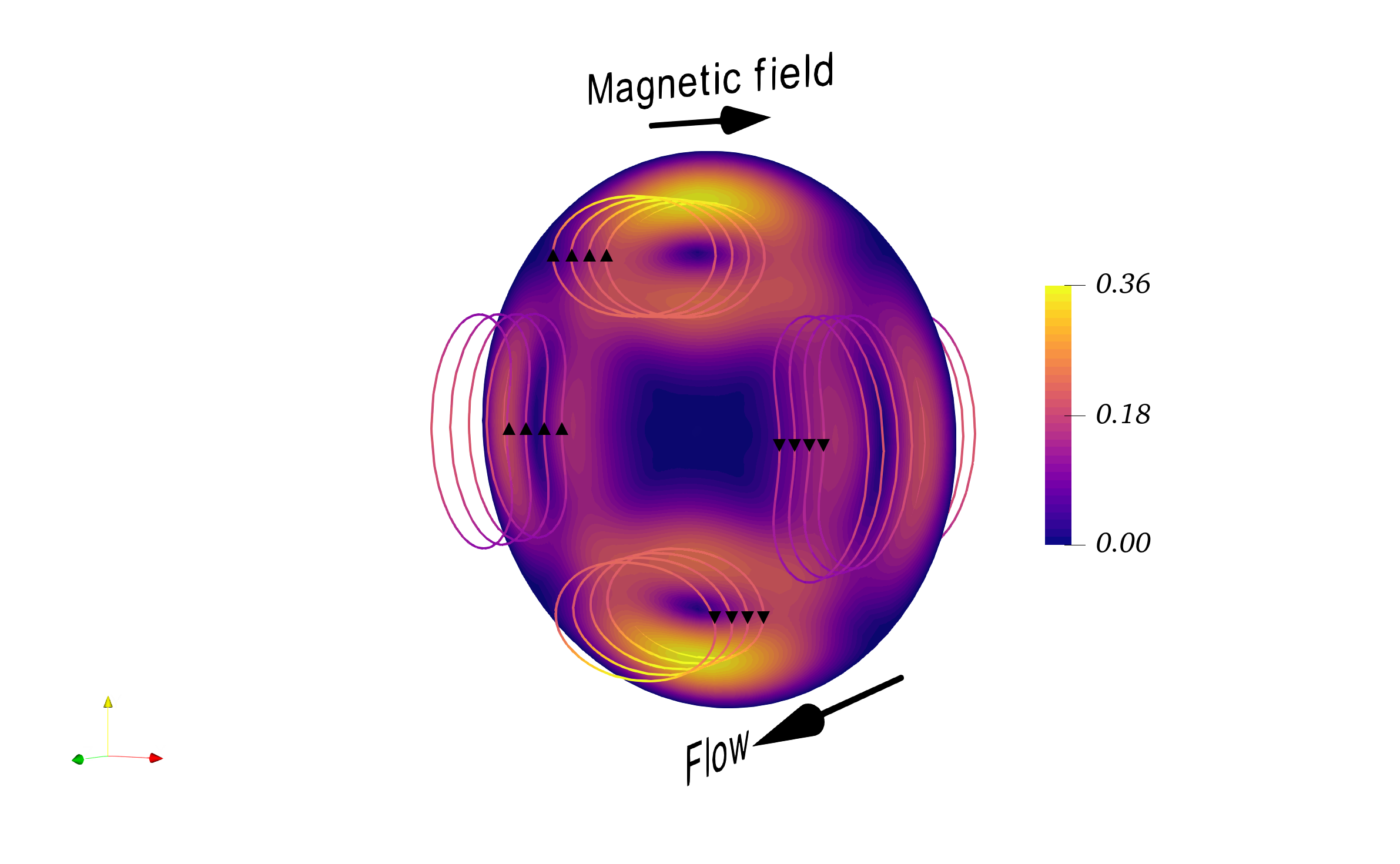}
  \end{subfigure}
  \begin{subfigure}[t]{0.4\textwidth}
    \centering
    \includegraphics[width=\textwidth,trim={30cm 2cm 12cm 1cm},clip]{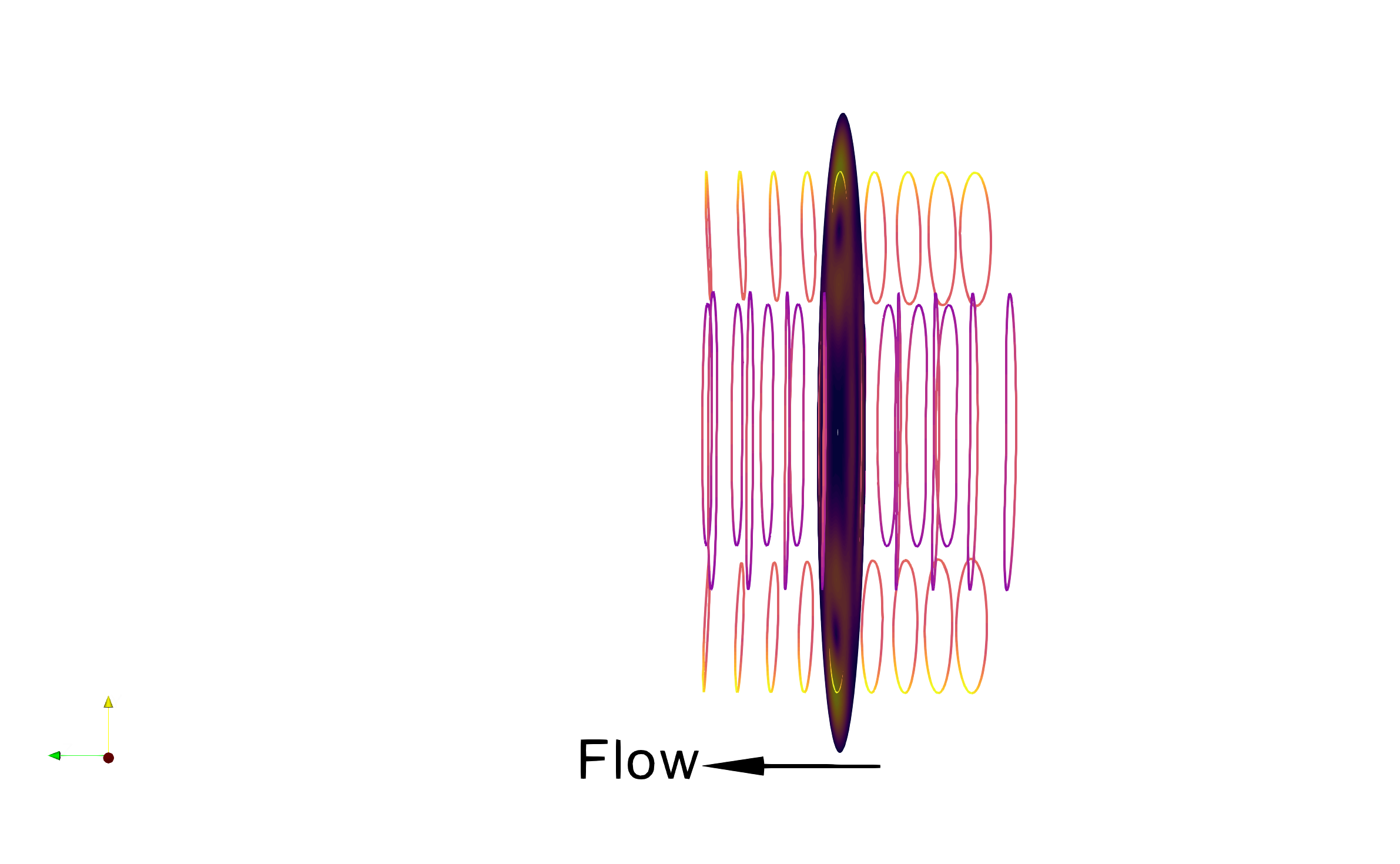}
  \end{subfigure}
  \caption{Contours of velocity magnitude and streamlines of the optimal perturbation at $t_{\textrm{opt}}=0$ for $Re=5000$ and $Ha=5$. The plots are shown for $\sigma_w=0$ (top) and $\sigma_w=\infty$ (bottom) and at $z = L_\alpha / 2$, where $L_\alpha$ is perturbation wavelength. In both cases, the perturbation consists in a quartet of streamwise vortices.}
  \label{figure:initial-streamlines-5-5000}
\end{figure}

\begin{figure}
  \centering
  \begin{subfigure}[t]{0.9\textwidth}
    \centering
    \includegraphics[width=\textwidth,trim={0 1.15cm 0 0},clip]{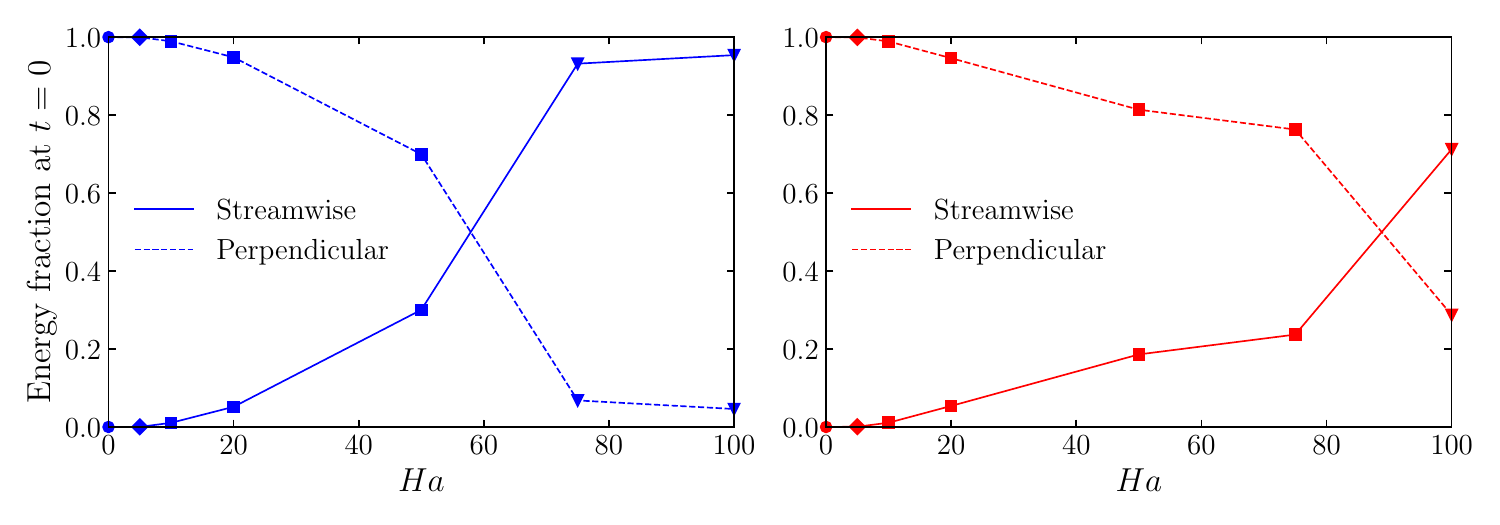}
  \end{subfigure}
  \vskip\baselineskip
  \begin{subfigure}[t]{0.9\textwidth}
    \centering
    \includegraphics[width=\textwidth,trim={0 0 0 0.25cm},clip]{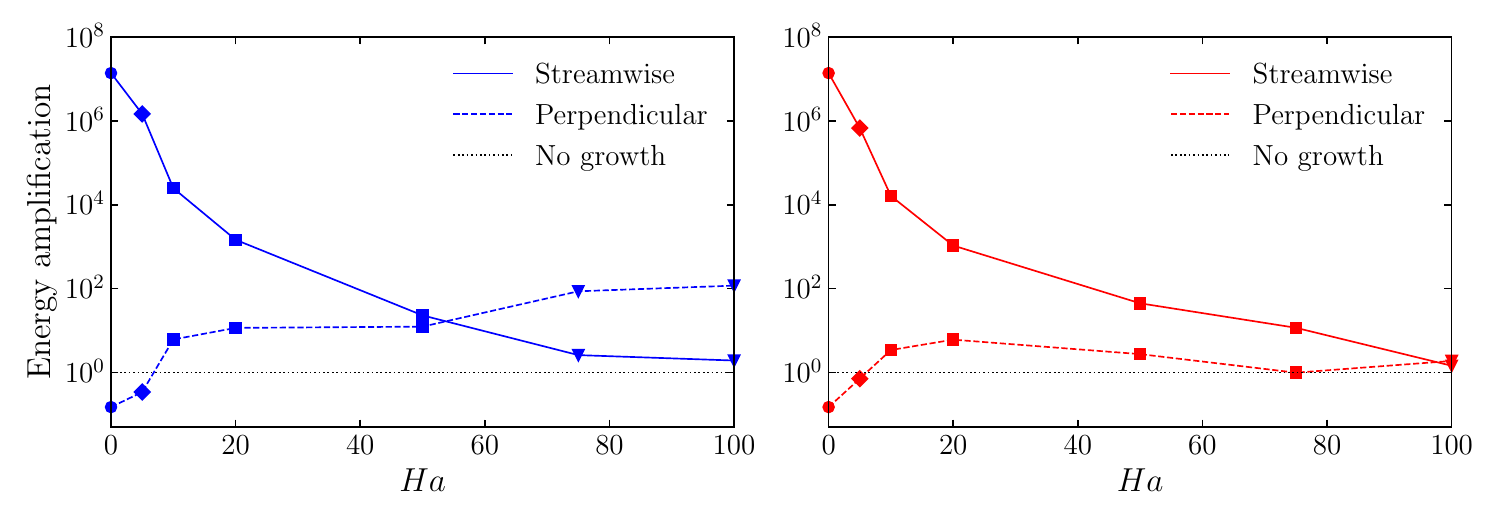}
  \end{subfigure}
\caption{Fraction of perturbation kinetic energy contained in the streamwise and perpendicular velocity components of the optimal perturbation at $t=0$ (top), and amplification of the streamwise and perpendicular velocity components at $t=t_{\textrm{opt}}$ (bottom) as functions of the Hartmann number for $Re=5000$. Results are shown for $\sigma_w=0$ ({\color{blue}blue}) and $\sigma_w=\infty$ ({\color{red}red}). The different markers represent the different topologies of optimal perturbations (see Fig.~\ref{figure:optimal-growth} for details).}
\label{figure:component-energy-gain-5000}
\end{figure}

For $3 \lesssim Ha \lesssim 10$, a second type of perturbation is globally optimal.
Considering, for example, $Ha=5$, we plot in Fig.~\ref{figure:initial-streamlines-5-5000} its streamlines and the corresponding contours of kinetic energy (for both $\sigma_w = 0$ and $\sigma_w = \infty$) at $t=0$.
This perturbation is still characterized by a weak streamwise dependence and takes the form of a quartet of streamwise vortices occupying a large fraction of the pipe's cross-section. As a consequence, the streamwise velocity component of such a perturbation contains only 0.04\% (resp. 0.08\%) of total kinetic energy when the pipe's wall is perfectly insulating (resp. perfectly conducting) (see Fig.~\ref{figure:component-energy-gain-5000} (top)).
According to the classification adopted in Ref.~\cite{TatsumiYoshimura1990} for the case of the duct flow, this optimal perturbation is characterized by the two-plane symmetry of type II and is therefore polar symmetric.

\begin{figure}
  \centering
  \begin{subfigure}[t]{0.45\textwidth}
    \centering
    \includegraphics[width=\textwidth,trim={20cm 0.5cm 8cm 0},clip]{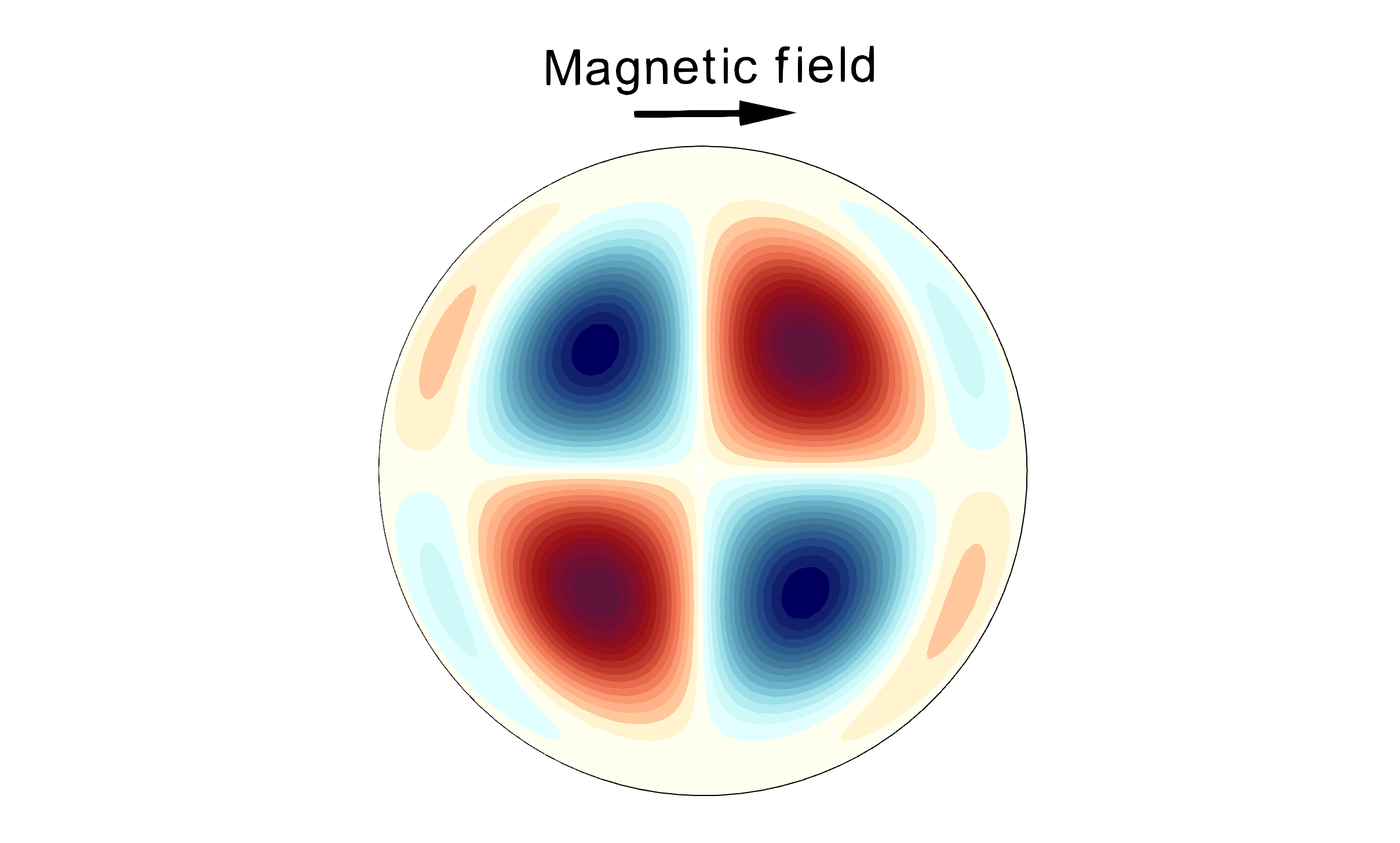}
  \end{subfigure}
  \begin{subfigure}[t]{0.45\textwidth}
    \centering
    \includegraphics[width=\textwidth,trim={20cm 0 8cm 0},clip]{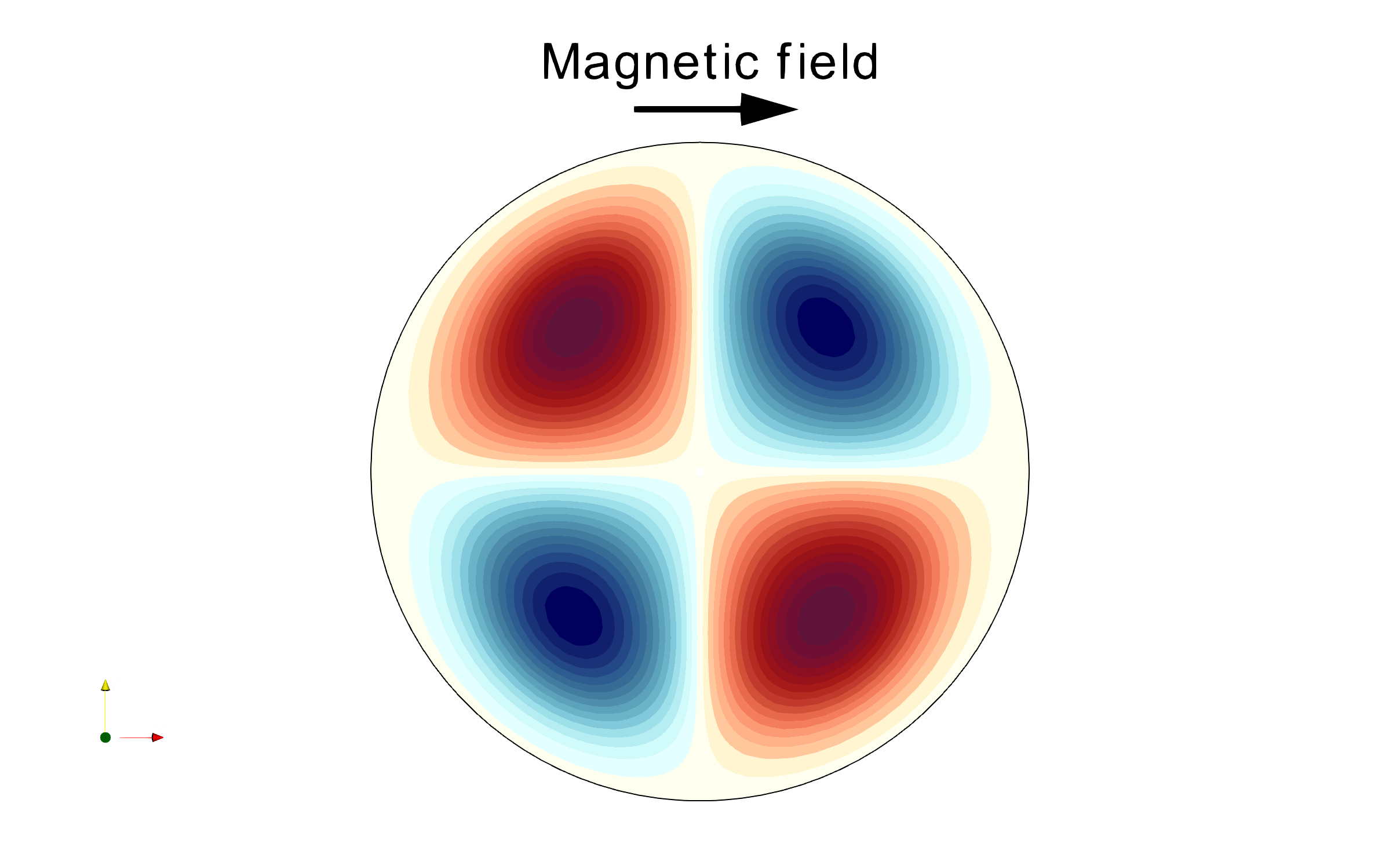}
  \end{subfigure}
  \caption{Contours of the streamwise velocity component $u_z$ of the optimal perturbation at $t = t_{\textrm{opt}}$ for $Re=5000$ and $Ha=5$. The plots are shown for $\sigma_w=0$ (left) and $\sigma_w=\infty$ (right) and $L_\alpha / 2$, where $L_\alpha$ is perturbation wavelength. The perturbation consists in high ({\color{red}red}) and low-speed ({\color{blue}blue}) nearly streamwise streaks.}
  \label{figure:evolved-cross-contours-5-5000}
\end{figure}

At the time of maximum amplification $t_{\textrm{opt}}$, this optimal perturbation evolves into a quartet of high and low-velocity longitudinal streaks with nearly 100\% of the total kinetic energy contained in the streamwise velocity component.
This is confirmed by examining the contours of $u_z$ plotted in Fig.~\ref{figure:evolved-cross-contours-5-5000} at the streamwise location $z = L_\alpha / 2$, where $L_\alpha$ denotes the perturbation wavelength.
In the perfectly insulating case, we also note the existence of another set of weak streaks in the Hartmann layers in opposite directions from their neighboring more intense streak. The amplification of the streamwise velocity originates from the lift-up effect in which high (resp. low) speed regions of the longitudinal base flow are advected by the original streamwise vortices toward the regions of low (resp. high) speed regions of the base flow. This is confirmed by examining in Fig.~\ref{figure:initial-streamlines-5-5000} the orientation of the streamlines and their location in the pipe's cross-section.
As can be seen from component-wise energy analysis of optimal modes at $t=t_{\textrm{opt}}$ in Fig.~\ref{figure:component-energy-gain-5000} (bottom), this mechanism is clearly dominant in the range of Hartmann numbers for which this perturbation is globally optimal. Finally, we note that this type of optimal perturbation exists in the hydrodynamic pipe flow but is only `locally' optimal for $t \lesssim 158$.
\subsubsection{Intermediate values of \textit{Ha}}
\begin{figure}
  \centering
  \begin{subfigure}[t]{0.8\textwidth}
      \centering
      \includegraphics[width=\textwidth,trim={5cm 10cm 10cm 5cm},clip]{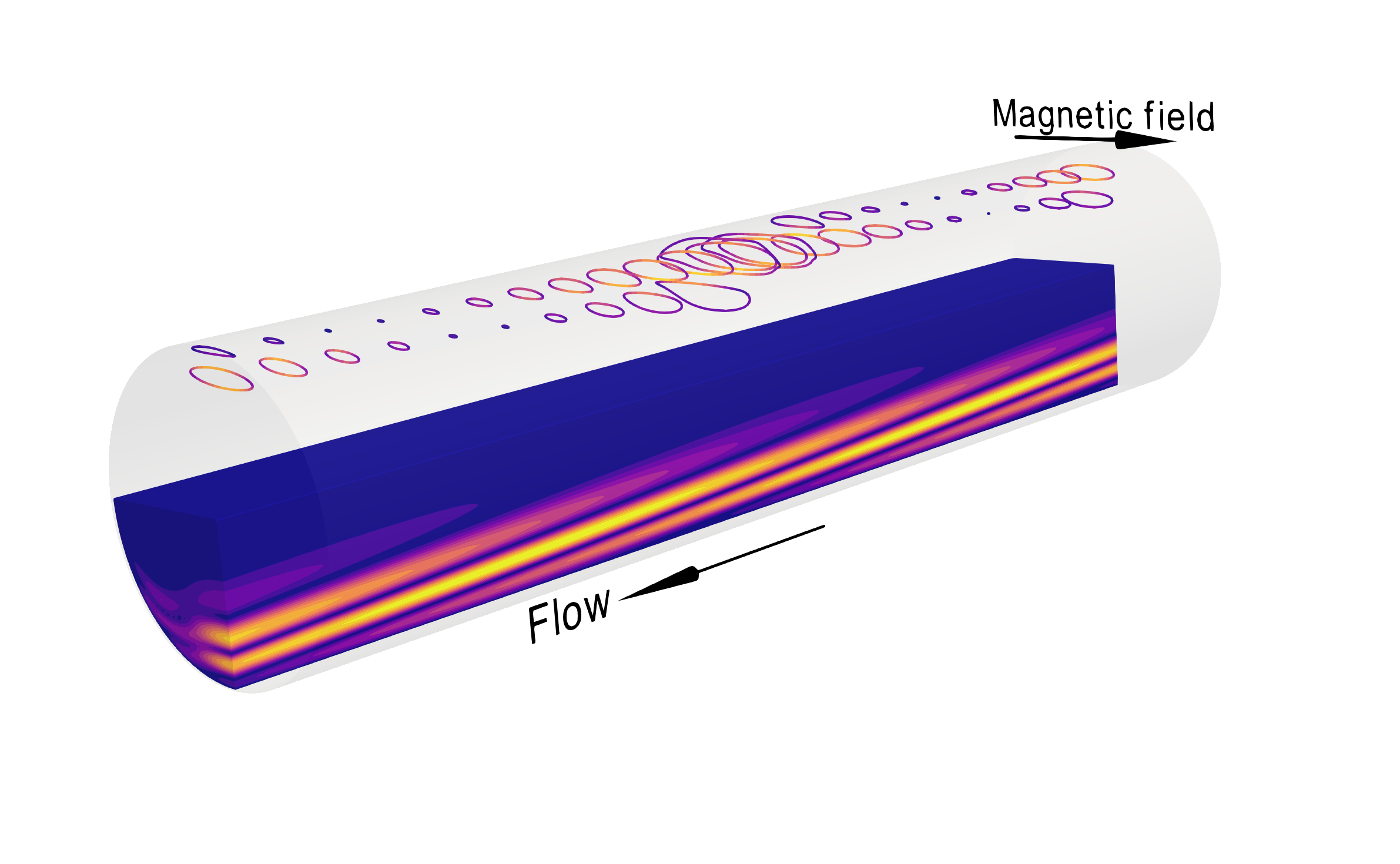}
  \end{subfigure}
  \vskip\baselineskip
  \begin{subfigure}[t]{0.8\textwidth}
      \centering
      \includegraphics[width=\textwidth,trim={10cm 7cm 13cm 10cm},clip]{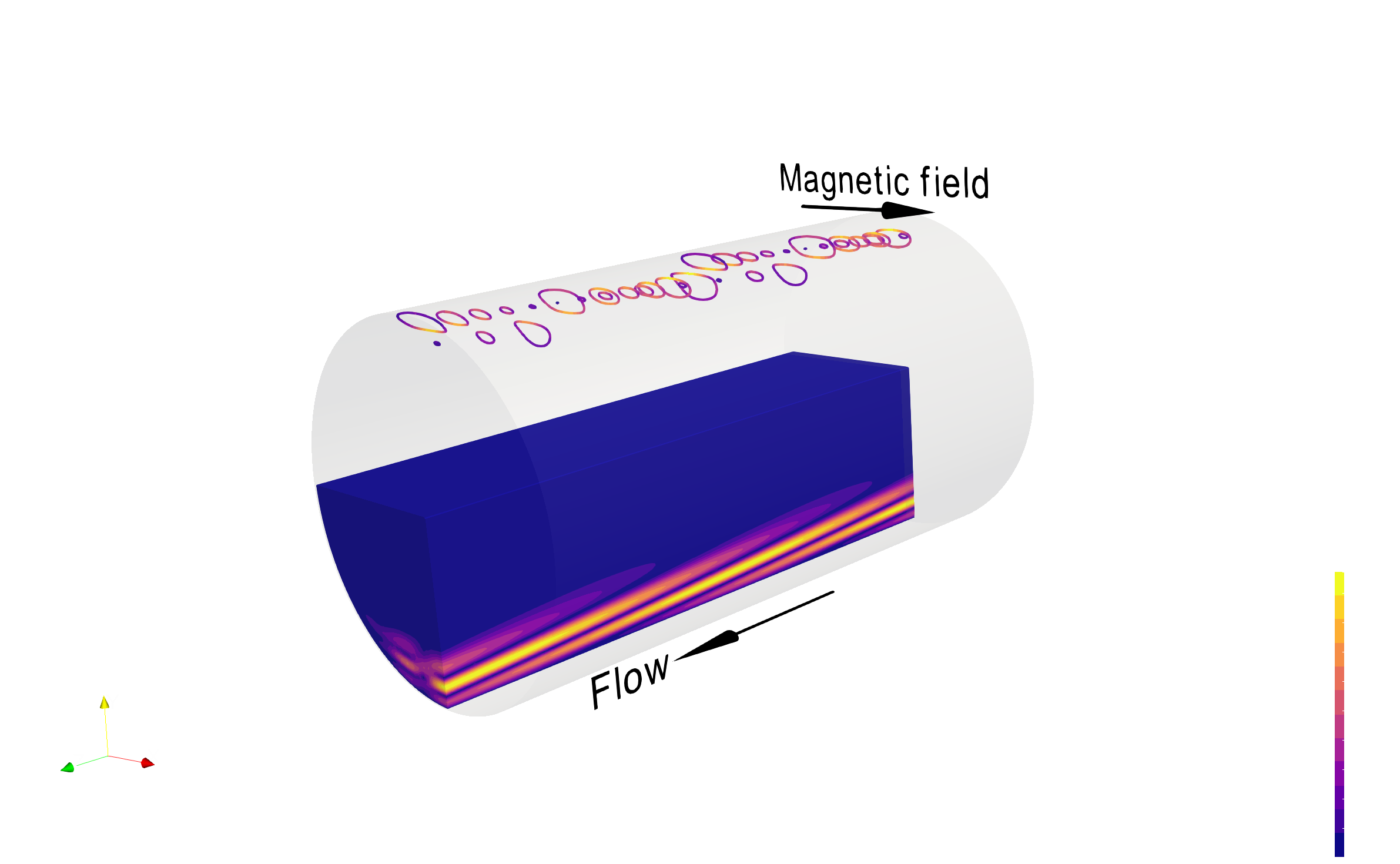}
  \end{subfigure}
  \caption{Contours of velocity magnitude and the streamlines of the optimal perturbation at $t=0$ for $Re=5000$ and $Ha=20$. The plots are shown for $\sigma_w=0$ (top) and $\sigma_w=\infty$ (bottom) and a single perturbation wavelength. The perturbation takes form of oblique vortices in the Roberts layers of the flow.}
  \label{figure:initial-streamlines-contours-20-5000}
\end{figure}

For $10 \lesssim Ha \lesssim 50$ in the perfectly insulating case and $10 \lesssim Ha \lesssim 75$ in the perfectly conducting case, we observe a third type of optimal perturbations. Notably, it is similar to that observed in the MHD duct flow with transverse magnetic field for moderate values of $Ha$ in Refs.~\cite{KrasnovZikanovRossiBoeck2010} and \cite{CassellsVoPotheratSheard2019}. Compared to the previously described optimal perturbation, it now possesses the two-plane symmetry of type IV according to the classification in Ref.~\cite{TatsumiYoshimura1990} and is therefore polar antisymmetric.
For $Ha=20$, we plot in Fig.~\ref{figure:initial-streamlines-contours-20-5000}
its streamlines and the corresponding contours of kinetic energy for $\sigma_w = 0$ and $\sigma_w = \infty$ at $t=0$.
In both cases, the motion takes form of stacked vortices located in the Roberts layers of the flow, elongated in the direction of the magnetic field and tilted `against' the mean flow in the streamwise direction.

Compared to the case at $Ha=5$, the fraction of kinetic energy contained in the streamwise velocity component is now approximately 5\% (see Fig.~\ref{figure:component-energy-gain-5000} (top)). For $Ha=50$ (no plots shown), the structure of the optimal perturbation is very similar. However, the tilting of the streamlines is more pronounced and the fraction of total energy contained in the streamwise velocity component at $t=0$ increases, reaching 30.1\% and 18.6\% in the perfectly insulating and perfectly conducting cases, respectively.

\begin{figure}
  \centering
  \begin{subfigure}[t]{0.45\textwidth}
    \centering
    \includegraphics[width=\textwidth,trim={15cm 3.5cm 28cm 12.5cm},clip]{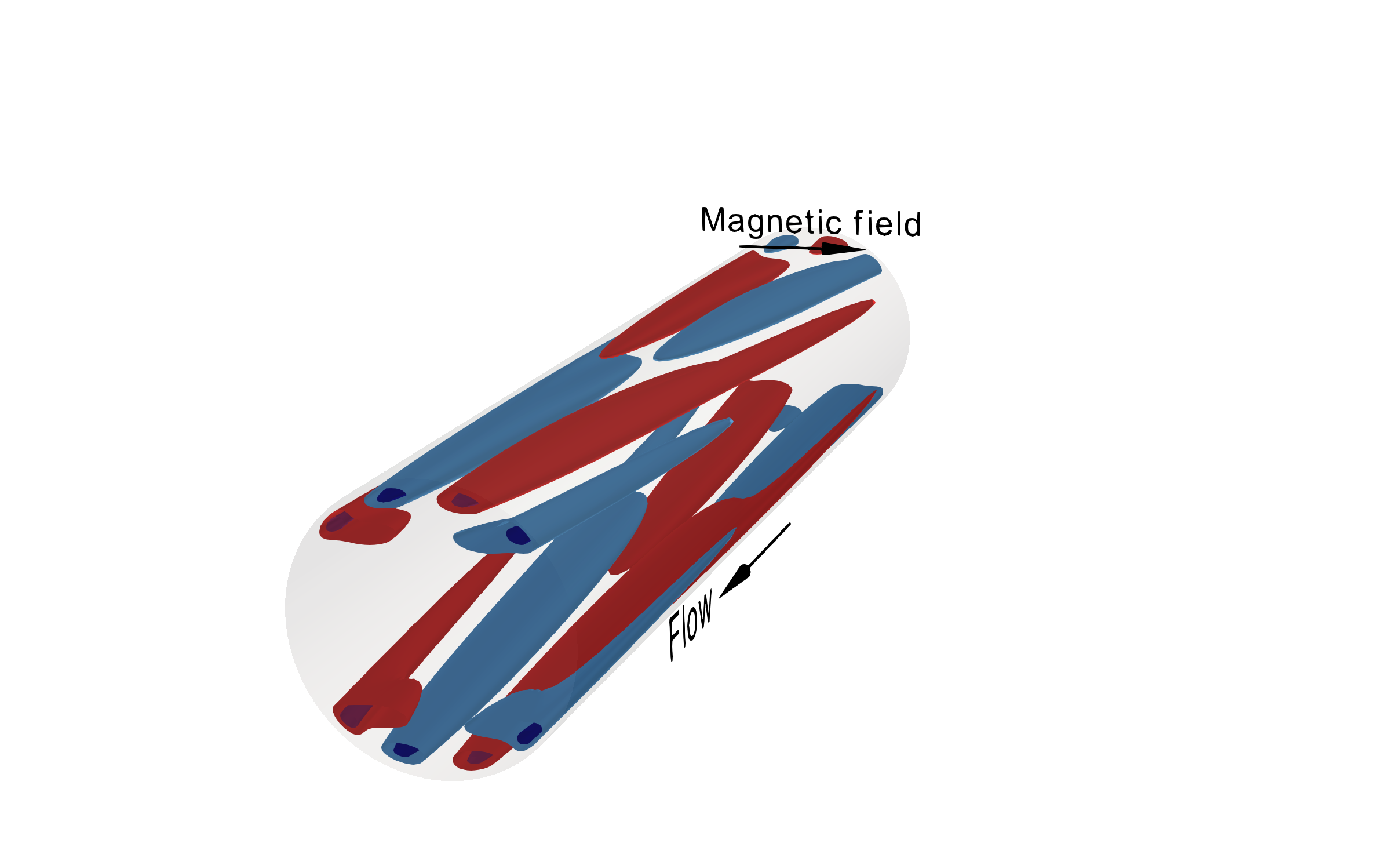}
    \put(-180,130){$\sigma_w = 0, t=5$}
  \end{subfigure}
  \begin{subfigure}[t]{0.45\textwidth}
    \centering
    \includegraphics[width=\textwidth,trim={15cm 3.5cm 27cm 11cm},clip]{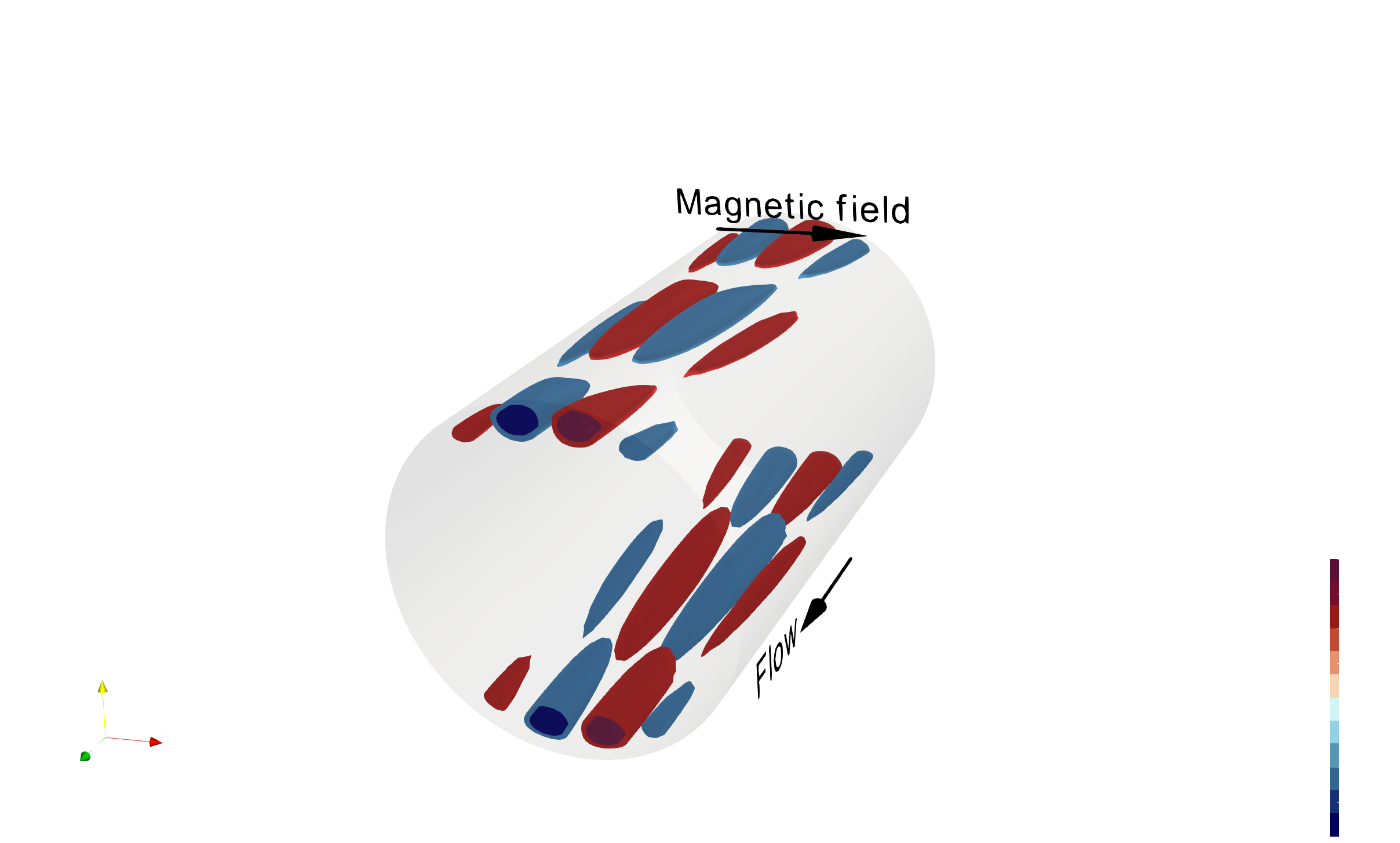}
    \put(-200,130){$\sigma_w = \infty, t=2.5$}
  \end{subfigure}
  \vskip\baselineskip
  \begin{subfigure}[t]{0.45\textwidth}
      \centering
      \includegraphics[width=\textwidth,trim={15cm 3.5cm 28cm 12.5cm},clip]{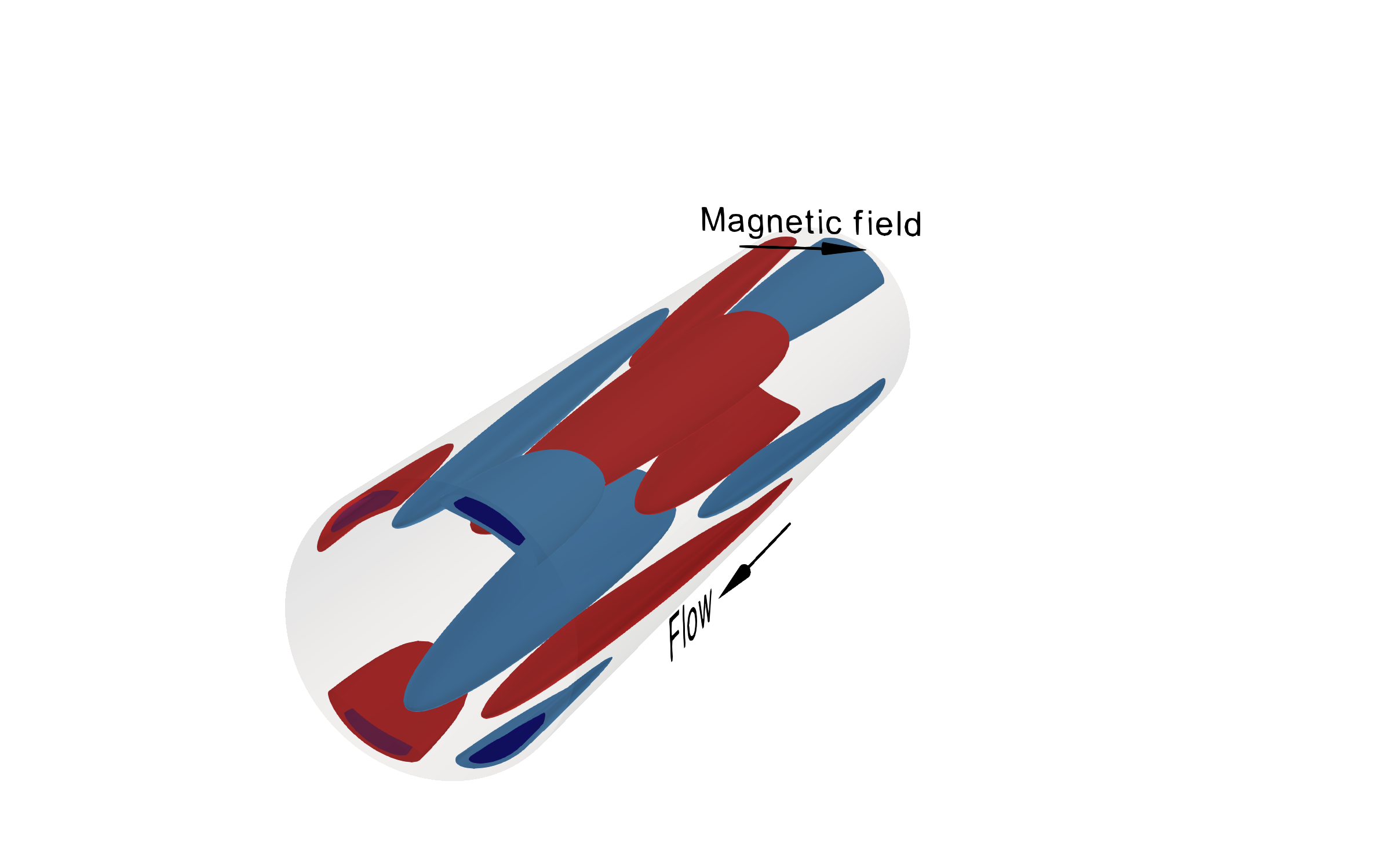}
      \put(-200,130){$\sigma_w = 0, t=t_\textrm{opt}$}
  \end{subfigure}
  \begin{subfigure}[t]{0.45\textwidth}
    \centering
    \includegraphics[width=\textwidth,trim={15cm 3.5cm 27cm 11cm},clip]{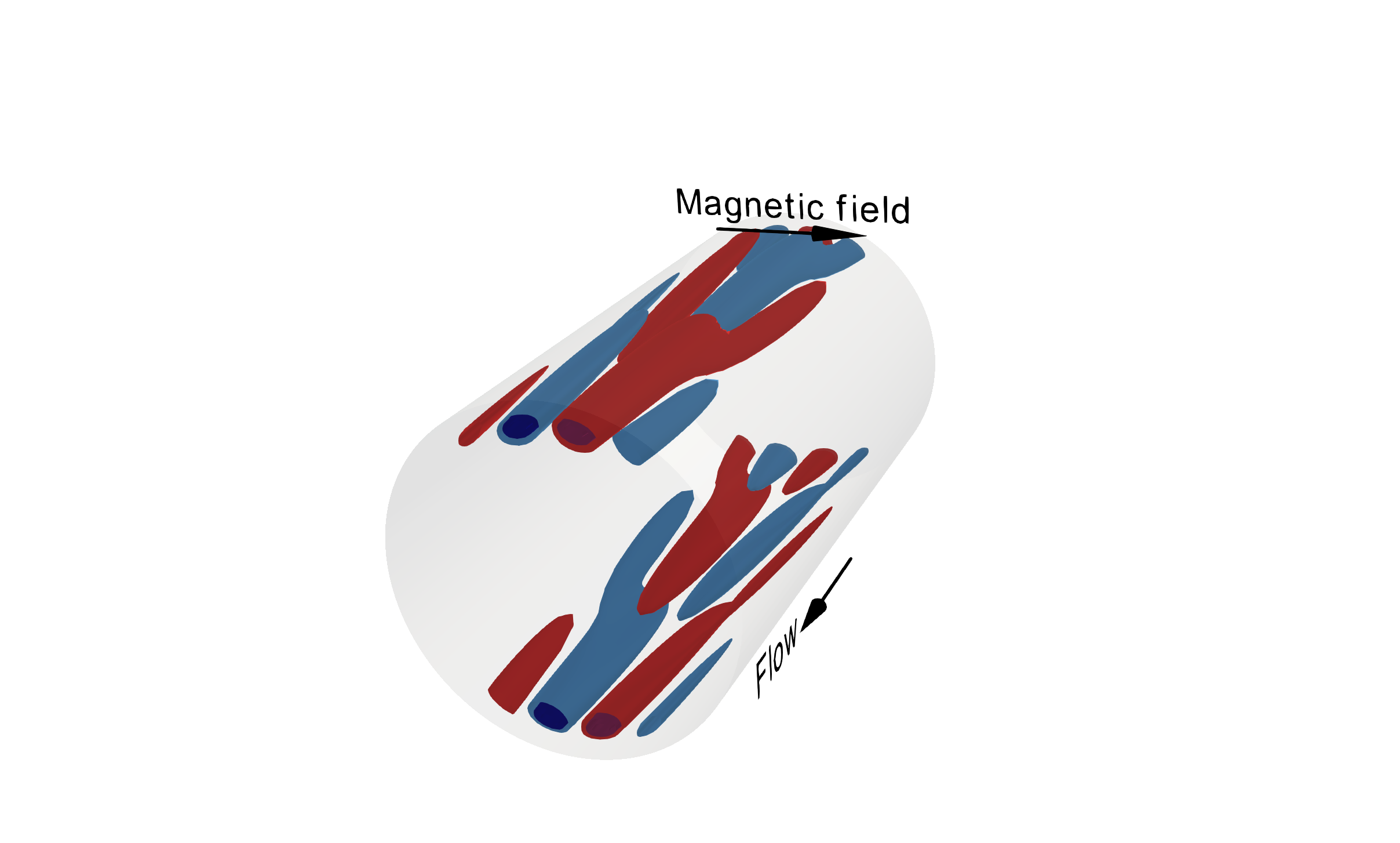}
    \put(-200,130){$\sigma_w = \infty, t=t_\textrm{opt}$}
  \end{subfigure}
  \caption{Isosurfaces of the streamwise velocity component of the optimal perturbation for $Re=5000$, $Ha=20$, $\sigma_w = 0$ (left column) and $\sigma_w=\infty$ (right column), for a single perturbation wavelength. The top row displays the initial creation of high ({\color{red}red}) and low-velocity ({\color{blue}blue}) streaks via the lift-up effect at $t=5$ and $t=2.5$, respectively. The bottom row displays the streaks when they have undergone secondary amplification and downstream tilting by the Orr-mechanism.}
  \label{figure:evolved-contours-20-5000}
\end{figure}

\begin{figure}
  \centering
  \includegraphics[width=0.6\textwidth,trim={0 0 0 0},clip]{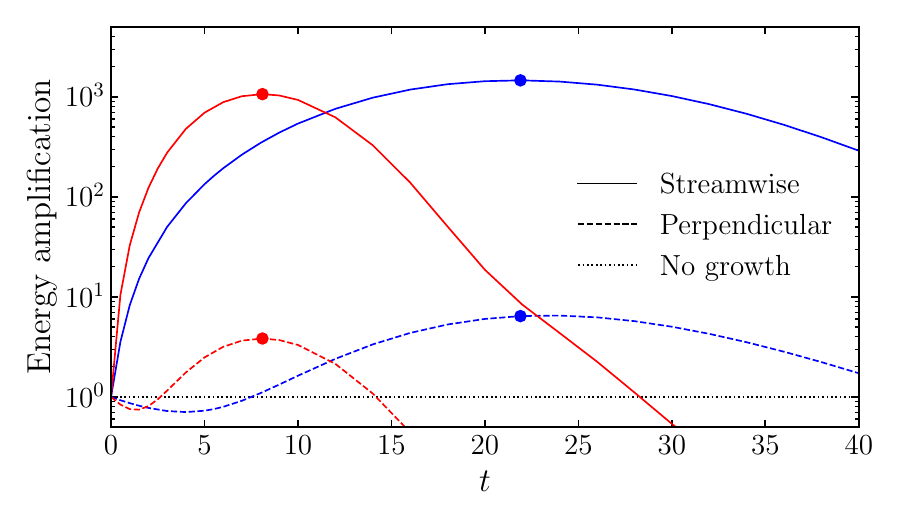}
  \caption{Amplification of the streamwise and perpendicular velocity components of the optimal perturbation as a function of time for $Re=5000$ and $Ha=20$. Results are shown for $\sigma_w=0$ ({\color{blue}blue}) and $\sigma_w=\infty$ ({\color{red}red}). The markers correspond to $t=t_{\textrm{opt}}$.}
  \label{figure:energy-evolution-by-components}
\end{figure}

At the time of maximum amplification $t_{\textrm{opt}}$, the perturbation consists in high and low-speed oblique streaks in the Roberts layers of the flow. The contribution of perpendicular velocity components in the total kinetic energy in the two cases then amounts to 7\% and 6\%, respectively. The amplification of this type of global perturbation occurs in two stages. At early times, the streamwise velocity component undergoes a rapid growth, while the perpendicular components decay. During this period, the original streamwise vortices are transformed into streaks through the lift-up mechanism, which remain tilted against the mean flow direction (see Fig.~\ref{figure:evolved-contours-20-5000} (top) for snapshots of the flow structures at $t=5$ for $\sigma_w = 0$ and $t=2.5$ for $\sigma_w = \infty$, respectively).
In the second stage, all velocity components are further amplified by the Orr-mechanism. As evident from Fig.~\ref{figure:evolved-contours-20-5000} (bottom), by the end of this subsequent phase, the streaks exhibit downstream tilting. This two-stage amplification of optimal perturbations is further illustrated in Fig.~\ref{figure:energy-evolution-by-components}, where we plot the evolution of streamwise and perpendicular kinetic energies (normalized with respect to their values at $t=0$) separately.
This highlights that after the initial amplification of the streamwise component of velocity by the lift-up effect, all components of velocity are then amplified similarly by the Orr-mechanism. In the case of hydrodynamic shear flows, this two-stage amplification scenario was first suggested in Ref.~\cite{Grossmann2000} and further discussed for boundary layers flows in Ref.~\cite{HoeffnerBrandtHenningson2005}. More recently, it was revisited in Ref.~\cite{JiaoHwangChernyshenko2021}, where the authors emphasized a transition scenario in which the Orr-mechanism precedes the lift-up effect by favoring optimal perturbations that maximize spanwise energy growth.

\subsubsection{High values of \textit{Ha}}
\begin{figure}
  \centering
\begin{subfigure}[t]{0.45\textwidth}
    \centering
    \includegraphics[width=\textwidth,trim={10cm 3.5cm 12cm 5cm},clip]{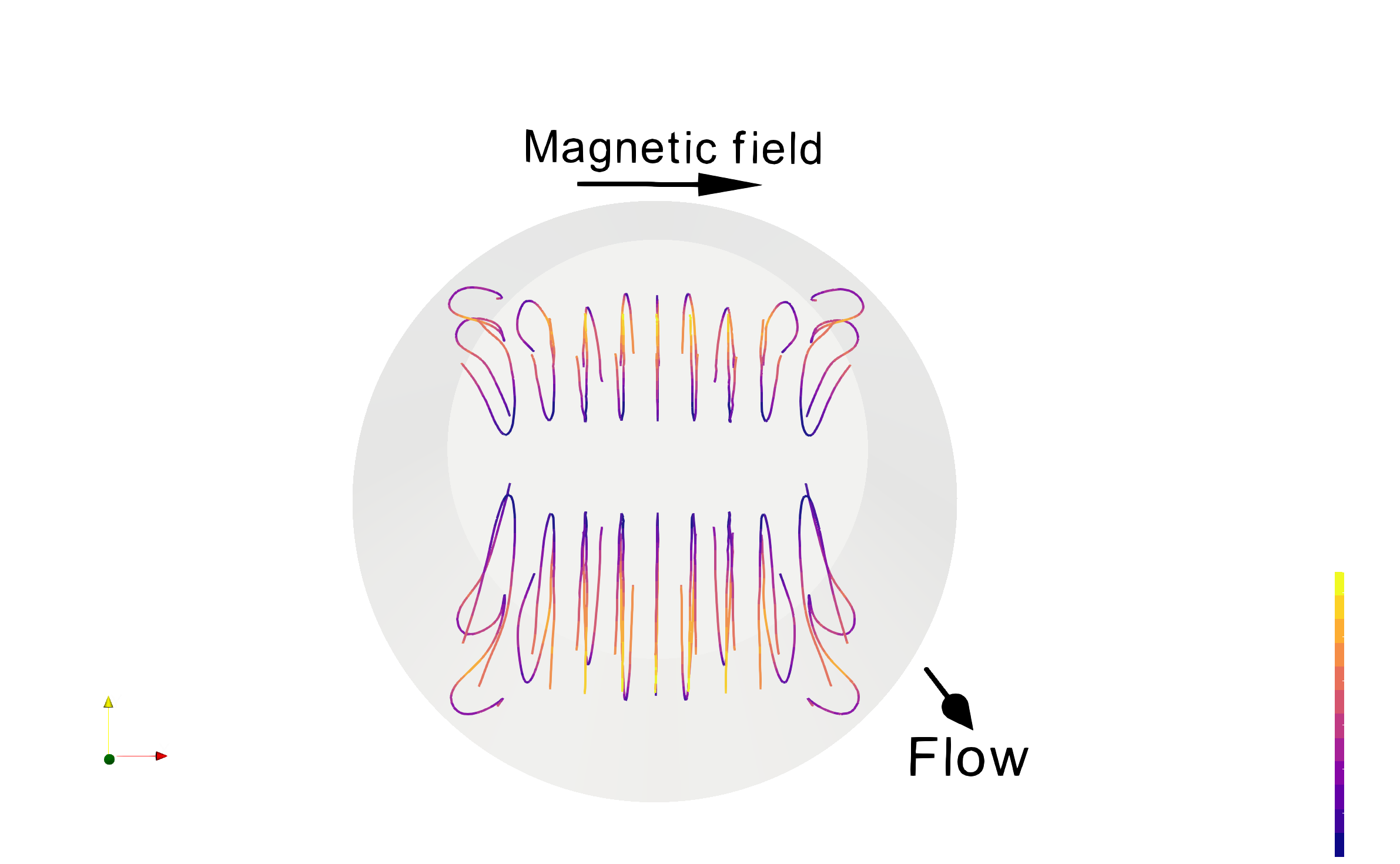}
    \put(-190,100){$t=0$}
\end{subfigure}
  \begin{subfigure}[t]{0.45\textwidth}
    \centering
    \includegraphics[width=\textwidth,trim={10cm 3.5cm 12cm 5cm},clip]{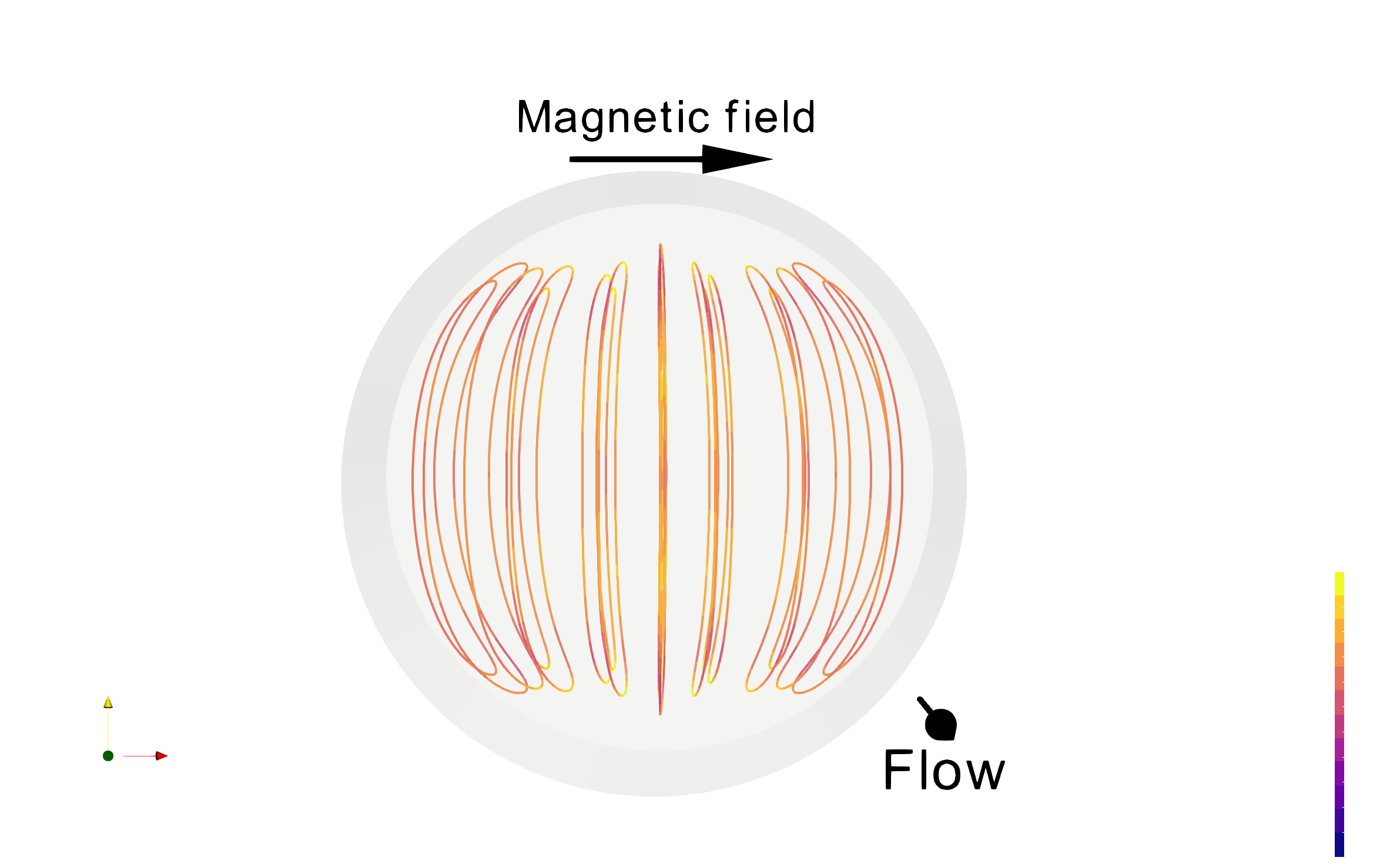}
    \put(-202,100){$t=t_{\textrm{opt}}$}
  \end{subfigure}
  \vskip\baselineskip
  \begin{subfigure}[t]{0.45\textwidth}
    \centering
    \includegraphics[width=\textwidth,trim={8.5cm 9cm 12cm 14.5cm},clip]{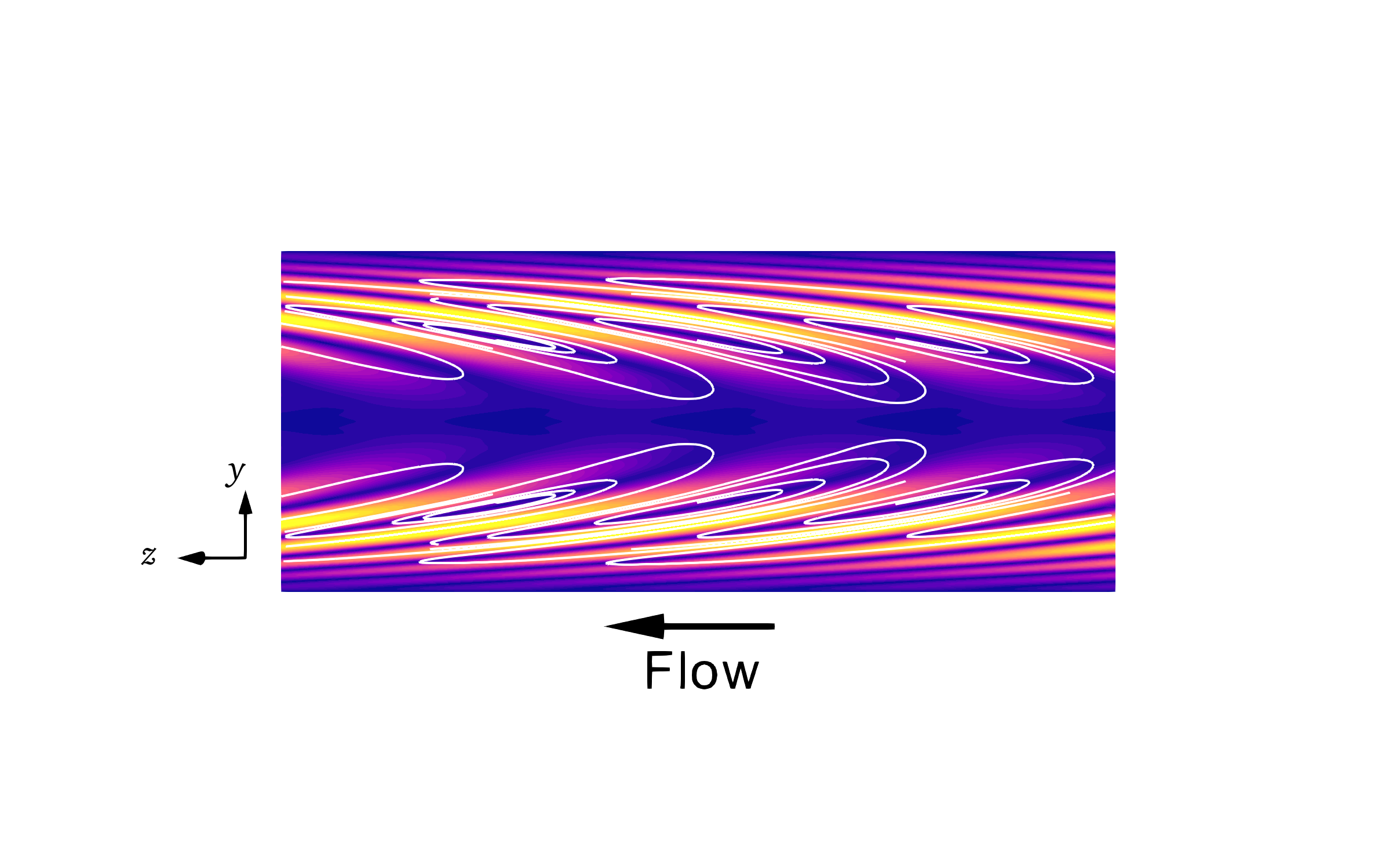}
    \put(-115,95){$t=0$}
  \end{subfigure}
  \begin{subfigure}[t]{0.45\textwidth}
      \centering
      \includegraphics[width=\textwidth,trim={8.5cm 9cm 12cm 14.5cm},clip]{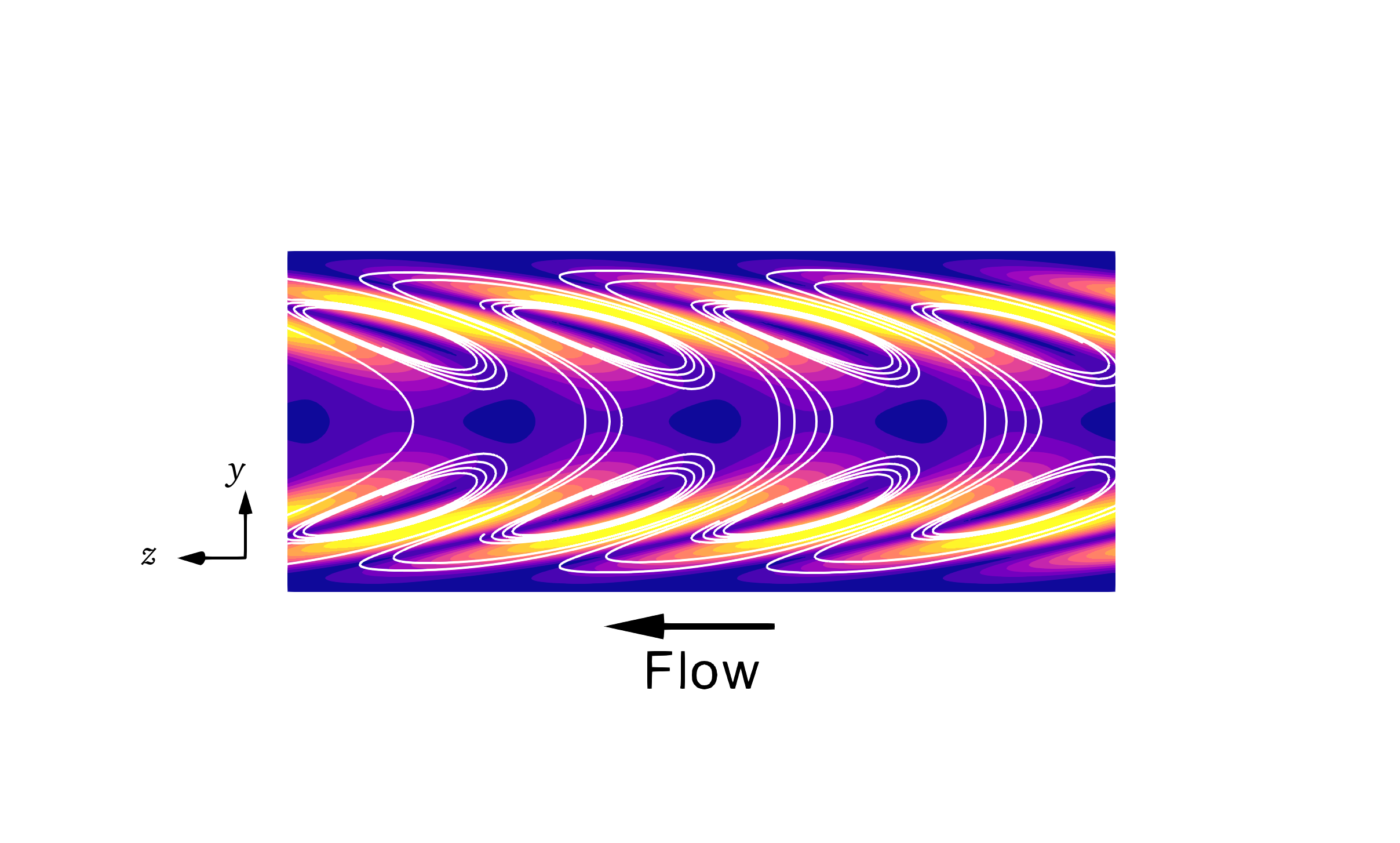}
      \put(-115,95){$t=4$}
  \end{subfigure}
  \vskip\baselineskip
  \begin{subfigure}[t]{0.45\textwidth}
    \centering
    \includegraphics[width=\textwidth,trim={8.5cm 9cm 12cm 14.5cm},clip]{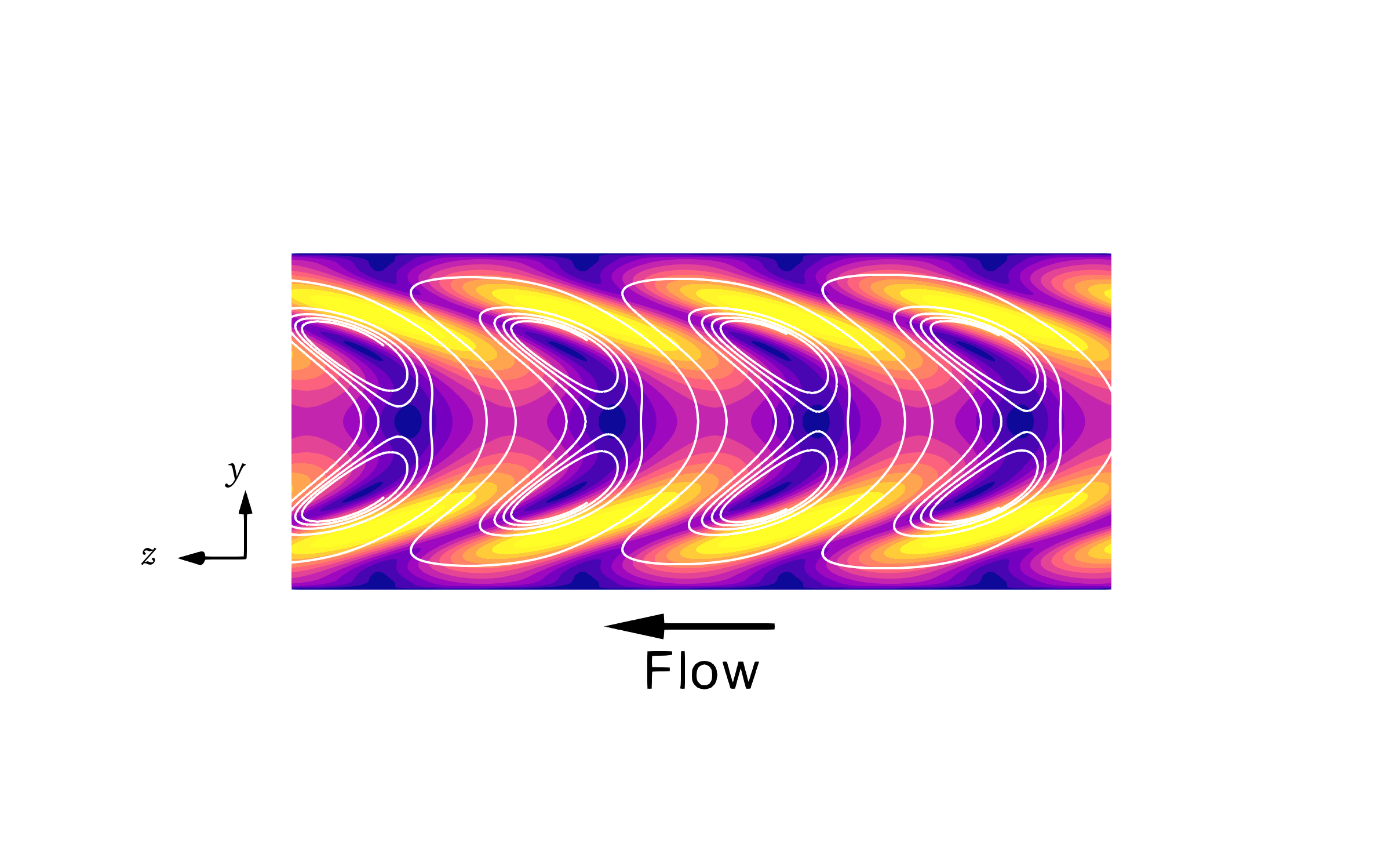}
    \put(-115,95){$t=6$}
  \end{subfigure}
  \begin{subfigure}[t]{0.45\textwidth}
    \centering
    \includegraphics[width=\textwidth,trim={8.5cm 9cm 12cm 14.5cm},clip]{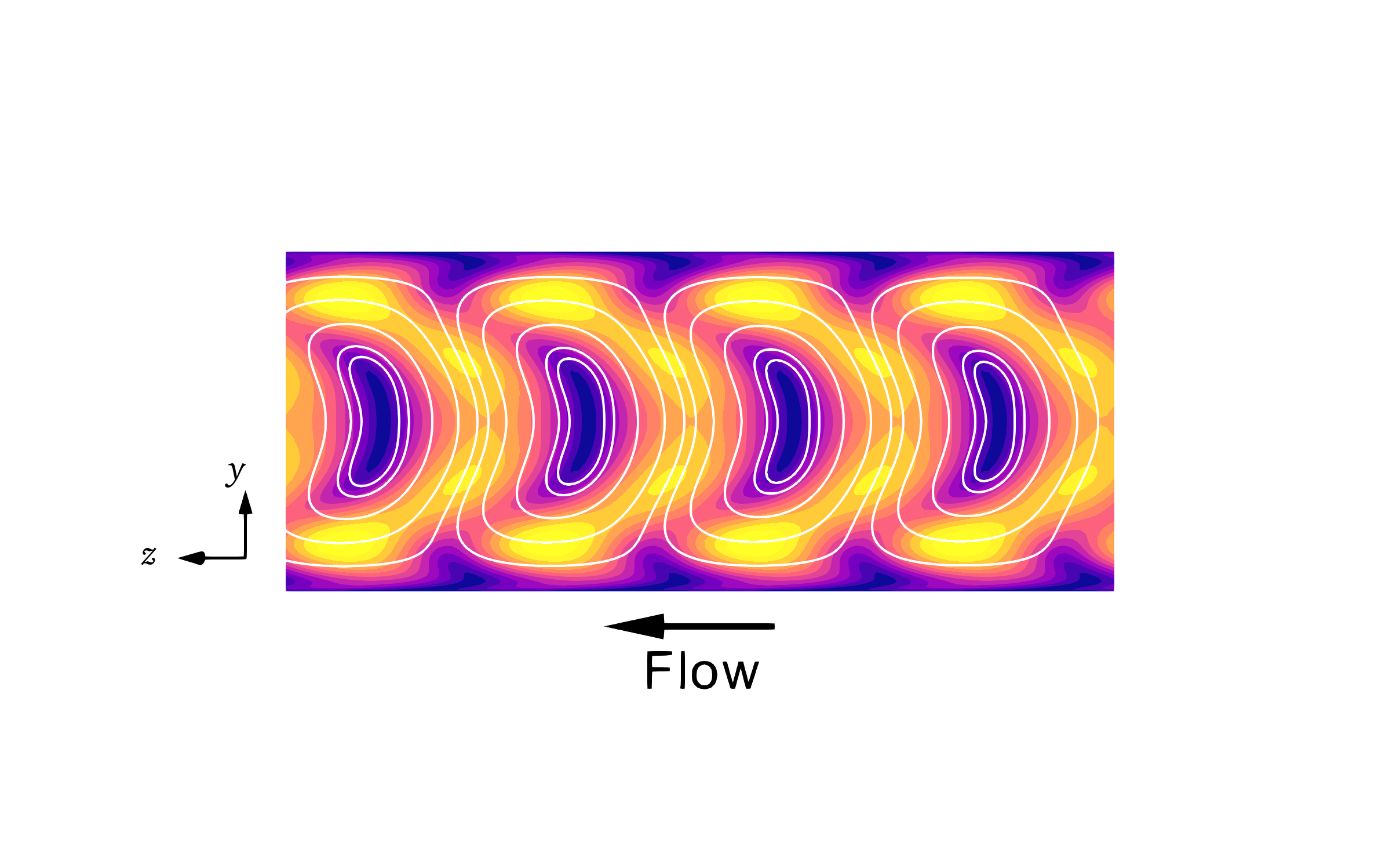}
    \put(-115,95){$t=t_{\textrm{opt}}$}
  \end{subfigure}
  \caption{Streamlines and contours of the velocity magnitude of the optimal perturbation for $Re=5000$, $Ha=100$ and $\sigma_w=0$. At $t=0$, the perturbation takes form of spanwise vortices in the Roberts layers, tilted upstream. Subplots at $t=4$ and $t=6$ illustrate the gradual merging of the streamlines. At $t=t_{\textrm{opt}}$, the spanwise vortices occupy the entire cross-section of the pipe.}
  \label{figure:streamlines-contours-100-5000}
\end{figure}

For $Ha\gtrsim 75$, a fourth type of optimal perturbation arises (we focus here solely on the perfectly insulating case, as in the perfectly conducting case, the transient growth is nearly fully damped by the magnetic field when this topology of the optimal mode can be observed). As an example, we display for $Ha=100$ the structure of the optimal perturbation in Fig.~\ref{figure:streamlines-contours-100-5000} for $t=0$ and $t=t_{\textrm{opt}}$.
At initial time, the motion consists in spanwise vortices, localized in the Roberts layers and strongly tilted against the mean flow. The perturbed flow is characterized by the two-plane symmetry of type I described in Ref.~\cite{TatsumiYoshimura1990}, and is therefore polar antisymmetric. For this value of $Ha$, 95\% of the total kinetic energy is contained in the streamwise velocity component. This type of ``quasi-2D" optimal perturbation is obviously favored under the action of a strong magnetic field promoting two-dimensional flow structures with no variations in its direction \cite{PotheratSommeriaMoreau2000}. By analogy with the MHD duct flow with strong transverse magnetic field, it is likely to persist at higher values of $Ha$ \cite{CassellsVoPotheratSheard2019}.

During their amplification, the spanwise rolls gradually assume a more upright position with respect to the mean flow. The streamlines, which were initially confined to separate Roberts layers, gradually merge at their tips, resulting in the formation of large-scale spanwise vortices that span the entire cross-section of the pipe and connect the Roberts layers. This amplification of the perpendicular components of the velocity field originates from the Orr-mechanism, which is the dominant mechanism of transient growth for $Ha \gtrsim 75$ in the perfectly insulating case (see Fig.~\ref{figure:component-energy-gain-5000}). In this flow regime, the initial perturbations and their time-evolved counterparts bear a strong resemblance to the optimal perturbations described in Ref.~\cite{Farrell1988} for the case of 2D plane Poiseuille flow. This similarity further asserts the quasi-2D nature of optimal perturbations for the MHD pipe flow for high Hartmann numbers.

\subsection{Nonlinear evolution of optimal perturbations}
\label{sec:nonlinear}
In this section, we study the nonlinear evolution of the optimal perturbations. The simulations are performed using the finite volume solver YALES2 \cite{MoureauDomingoVervisch2011} at Reynolds number $Re = 5000$ and for Hartmann numbers $Ha=5, 10, 20$. The solver uses a fourth-order node-based finite volume spatial discretization and the TRK4 fourth-order scheme for time advancement \cite{Kraushaar2011}. The Lorentz force appearing in Eq.~\eqref{eq:mhd-equations-2} is computed using the electric potential formulation as described in Ref.~\cite{Vantieghem2011}.
For each simulation, the length of the computational domain is $L_z = 4\pi/\alpha_{\mathrm{opt}}$, so that it contains two wavelengths of the optimal perturbation.
For $Ha=5, 10, 20$, the computational grids are fully unstructured and contain $10^8$, $2.2\cdot 10^7$ and $1.3\cdot 10^7$ grid nodes, respectively.
The results reported below mainly concern the amplitude of perturbations required to trigger a transition to turbulence, as well as the qualitative behavior of the flow structures before a fine-scale turbulent state is sustained. Using a grid convergence analysis, we have verified that these features are not affected when the resolution is increased further. As an additional validation of the numerical method, we confirmed that the nonlinear simulations matched the linear solutions for sufficiently small perturbation amplitudes.

The initial condition $\boldsymbol{u}_0$ for the direct numerical simulation is given by:
\begin{equation}\label{eq:dns-init-cond-no-noise}
  \boldsymbol{u}_0 = \boldsymbol{U} + \epsilon\boldsymbol{u}_\textrm{opt},
\end{equation}
where $\boldsymbol{U}$ is the two-dimensional stationary base flow, $\boldsymbol{u}_\textrm{opt}$ is the optimal perturbation and $\epsilon$ is the amplitude of the perturbation. The latter is chosen in such a way that at $t=0$, the fraction of kinetic energy of the fluctuating field with respect to the mean flow, $e(t) = E_{\boldsymbol{u}'(t) } / E_{\boldsymbol{U}}$, is a given value $e_0$, varied in decades from $10^{-5}$ to $10^{-2}$. In certain cases discussed further below, we introduce weak three-dimensional noise to the initial condition~\eqref{eq:dns-init-cond-no-noise}. The amplitude of the noise is chosen such that its kinetic energy is 5\% of the perturbation kinetic energy.

First, we consider the evolution of $e(t)$ when the fluctuating field is set to the global optimal perturbations for $Ha=5, 10$ and 20.
\begin{figure}
  \centering
  \begin{subfigure}[t]{0.45\textwidth}
      \centering
      \includegraphics[width=\textwidth,trim={0 1cm 0 0},clip]{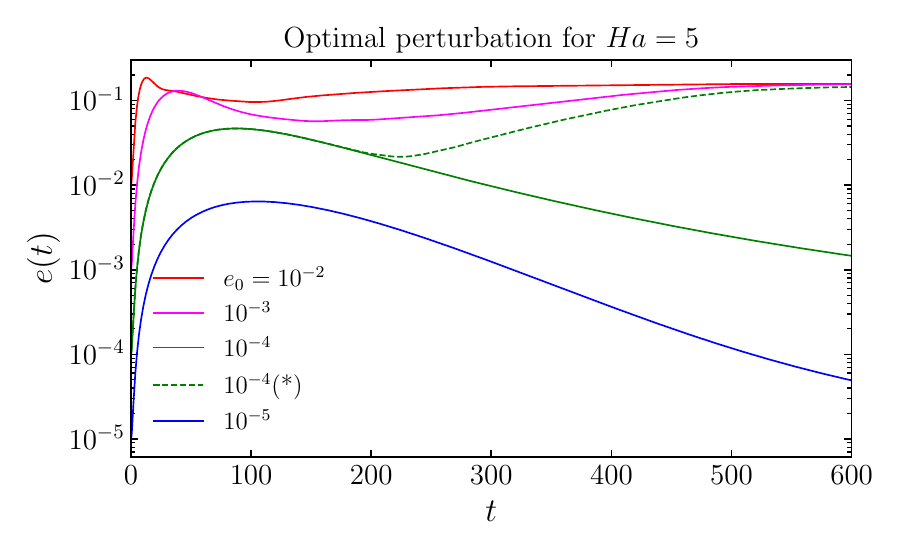}
  \end{subfigure}
  \begin{subfigure}[t]{0.4175\textwidth}
      \centering
      \includegraphics[width=\textwidth,trim={1.1cm 1cm 0 0},clip]{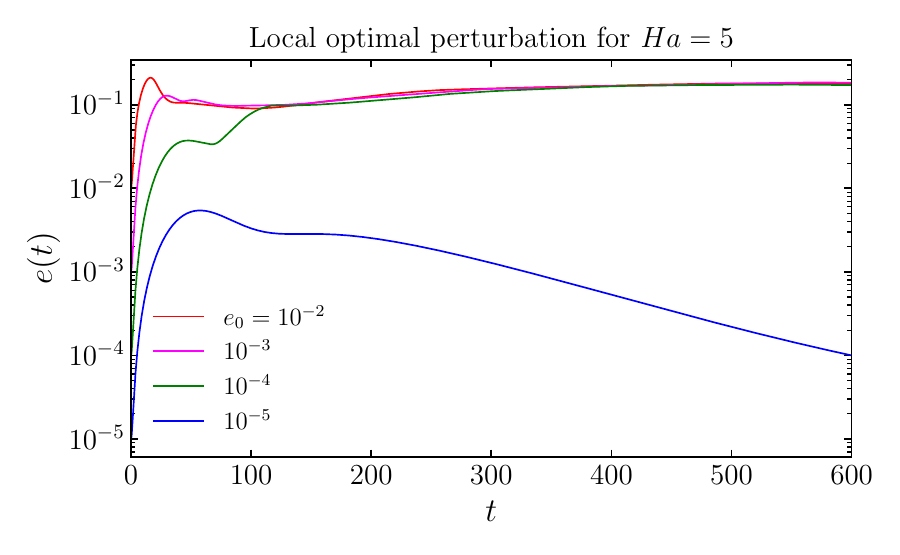}
  \end{subfigure}
  \vskip\baselineskip
  \begin{subfigure}[t]{0.45\textwidth}
    \centering
    \includegraphics[width=\textwidth,trim={0 0 0 0},clip]{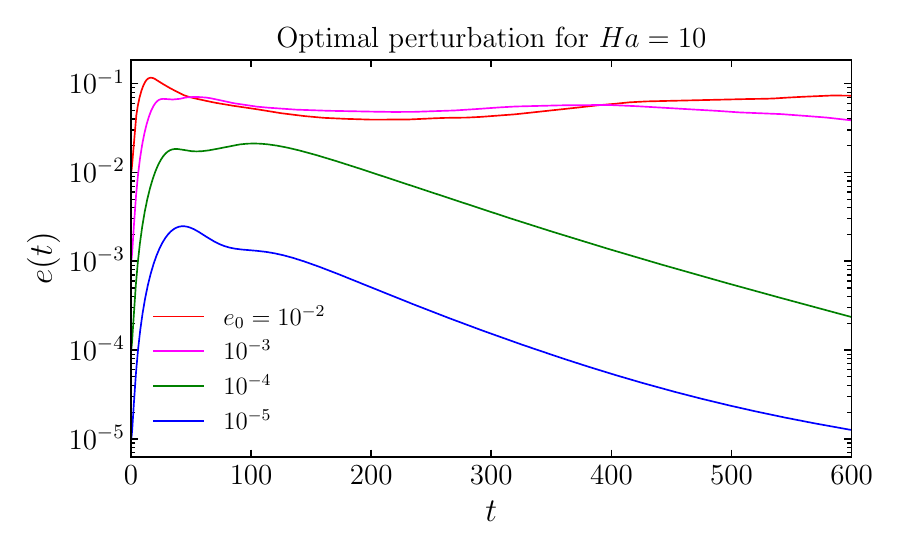}
  \end{subfigure}
  \begin{subfigure}[t]{0.4175\textwidth}
    \centering
    \includegraphics[width=\textwidth,trim={1.1cm 0 0 0},clip]{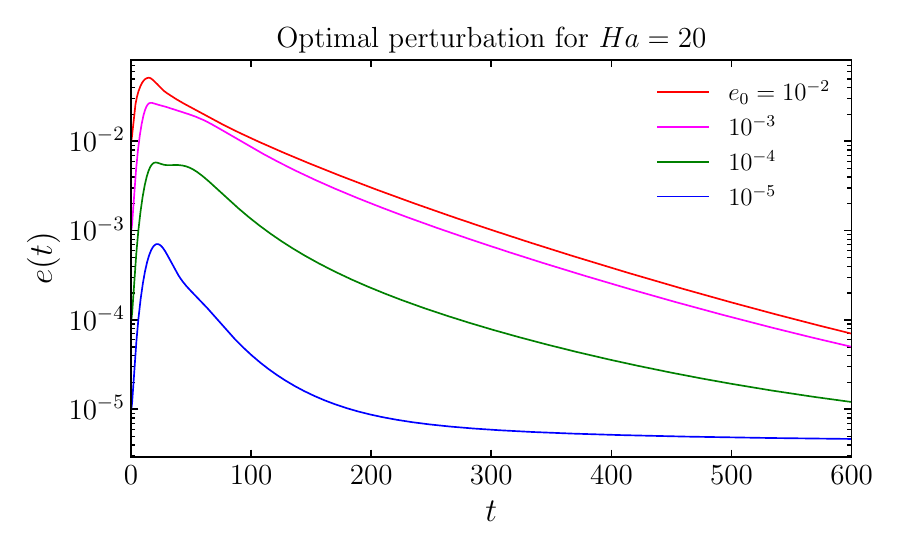}
  \end{subfigure}
  \caption{The nonlinear evolution of perturbation kinetic energy for $Re = 5000$ and $\sigma_w = 0$. Results are shown for the optimal perturbation at (top-left) $Ha = 5$, (bottom-left) $Ha = 10$ and (bottom-right) $Ha = 20$, and (top-right) the local optimal perturbation for $t=55$ at $Ha=5$.}
  \label{figure:nonlinear-energy-evolution-5000-0}
\end{figure}
The respective $e(t)$ curves for $Ha=5$ are displayed in Fig.~\ref{figure:nonlinear-energy-evolution-5000-0} (top-left). We observe that without the addition of random 3D noise, the transition to a sustained turbulent state occurs for $e_0 \gtrsim 10^{-3}$. With random noise added at $t=0$, this threshold is lowered to $e_0 \gtrsim 10^{-4}$. The time evolution of $e$ for cases corresponding to the global optimal perturbation at $Ha=10$ are shown in Fig.~\ref{figure:nonlinear-energy-evolution-5000-0} (bottom-left). In this case, the laminar-turbulent transition is initiated for $e_0 \gtrsim 2 \cdot 10^{-4}$, which is smaller than the value needed for the case $Ha=5$ (without 3D noise). Finally, as can be seen from Fig.~\ref{figure:nonlinear-energy-evolution-5000-0} (bottom-right), for $Ha=20$, the disturbances eventually decay for all values of $e_0$ considered, even with the addition of 3D noise.

Comparing Hartmann numbers $Ha=5$ and $Ha=10$, it appears that the former necessitates a larger initial fraction of the kinetic energy of the fluctuating field for its global optimal perturbation to trigger a transition to turbulence, despite experiencing less Joule damping. Since adding 3D noise to the initial condition for $Ha=5$ lowers the threshold for the transition, this behavior is likely due to the lack of three-dimensionality of the global optimal perturbation. To test this hypothesis, we have conducted simulations for $Ha=5$ using as the initial condition the (local) optimal perturbation at $t=55$. The amplification of this perturbation reaches 88\% of the global optimal amplification, but its topology is similar to that of the optimal perturbation observed for $10 \lesssim Ha \lesssim 50$ in the perfectly insulating case and $10 \lesssim Ha \lesssim 75$ in the perfectly conducting case. As such, its three-dimensionality is significantly more pronounced compared to the global optimal perturbation for the same Hartmann number. In Fig.~\ref{figure:nonlinear-energy-evolution-5000-0} (top-right), we display the time evolution of $e$ corresponding to this non-globally optimal perturbation. Contrary to the case of
the optimal perturbation, transition to turbulence can now be initiated for $e_0 \gtrsim 10^{-4}$, without the addition of random 3D noise. These observations reinforce the fact that while the optimal disturbance achieves the maximum linear amplification, a less amplified disturbance might be a more effective candidate for triggering nonlinear interactions.

\begin{figure}
  \centering
  \begin{subfigure}[t]{0.45\textwidth}
    \centering
    \includegraphics[width=\textwidth,trim={2cm 13cm 18cm 18cm},clip]{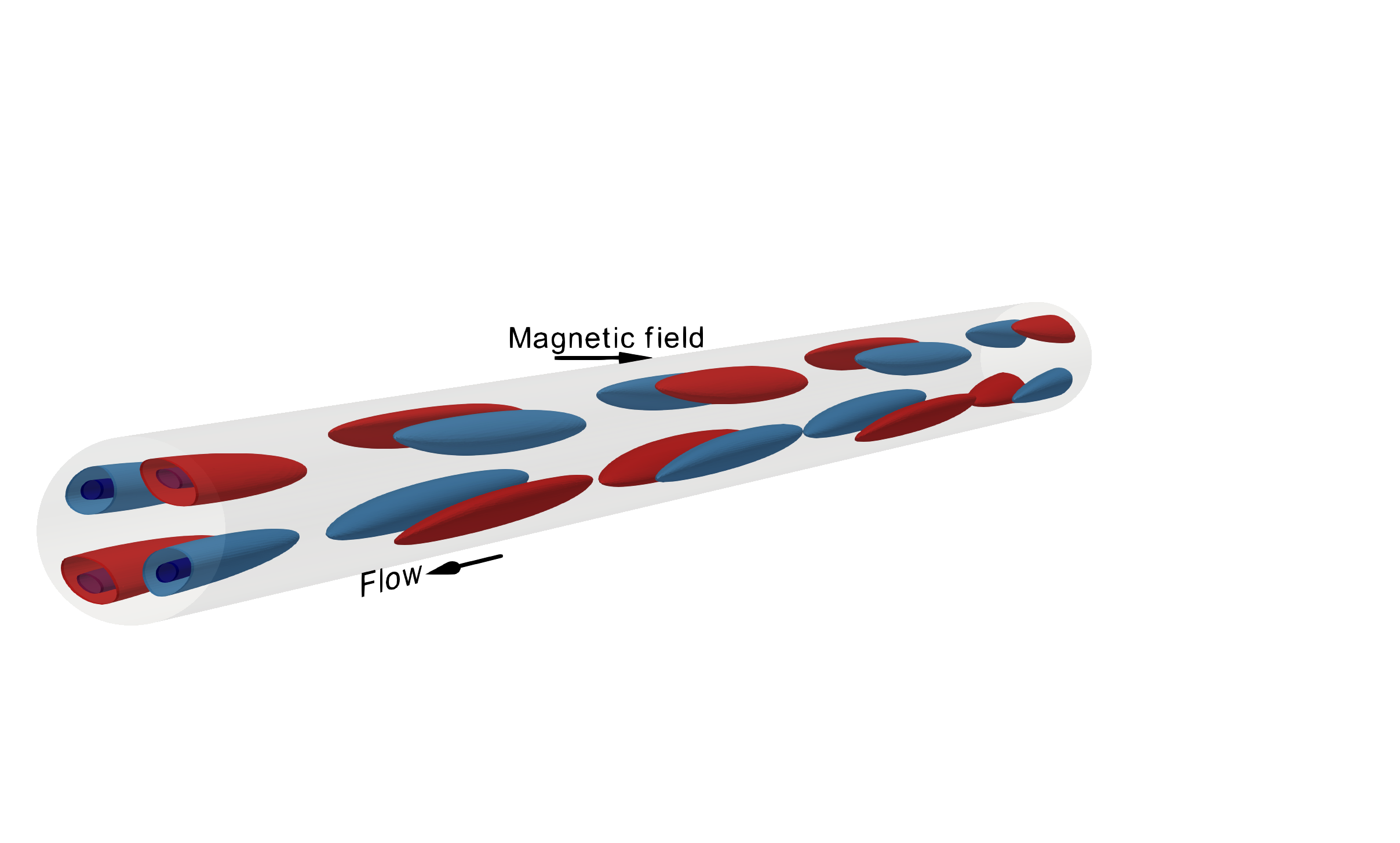}
    \put(-170,50){$t=3$}
  \end{subfigure}
  \begin{subfigure}[t]{0.45\textwidth}
    \centering
    \includegraphics[width=\textwidth,trim={2cm 13cm 18cm 18cm},clip]{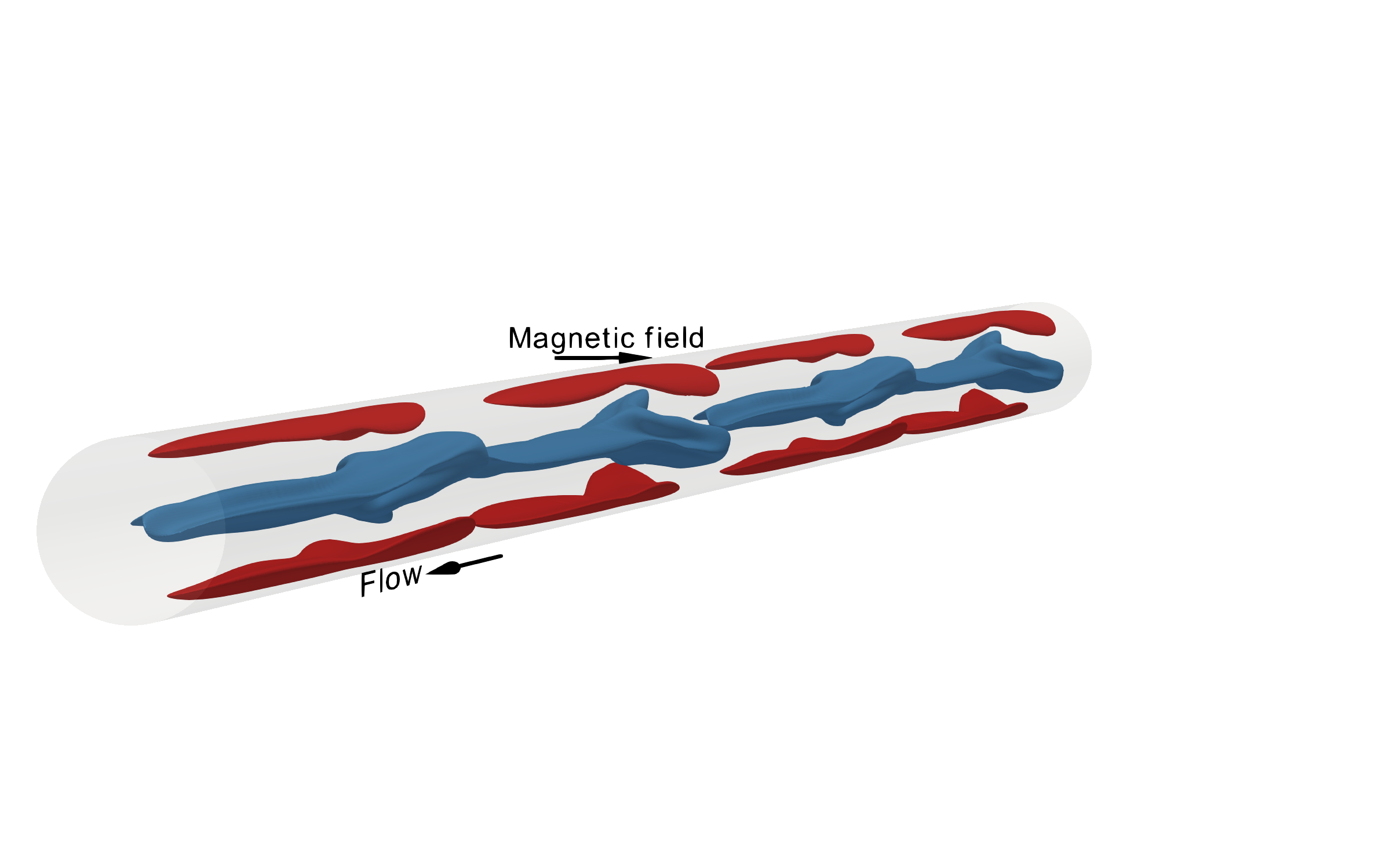}
    \put(-170,50){$t=t_\textrm{opt}$}
  \end{subfigure}
\vskip\baselineskip
  \begin{subfigure}[t]{0.45\textwidth}
    \centering
    \includegraphics[width=\textwidth,trim={2cm 13cm 18cm 18cm},clip]{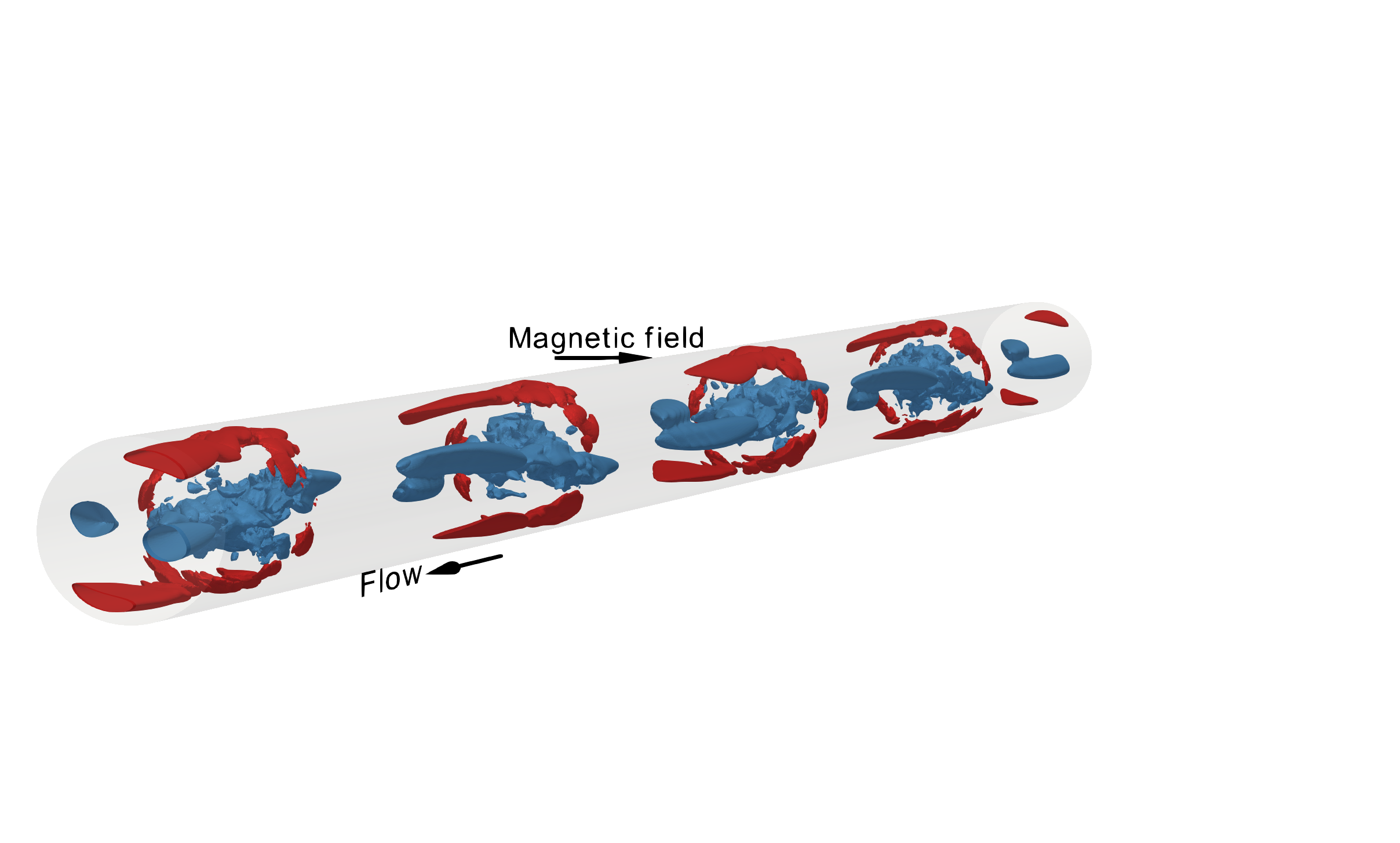}
    \put(-170,53){$t=2t_\textrm{opt}$}
  \end{subfigure}
  \begin{subfigure}[t]{0.45\textwidth}
    \centering
    \includegraphics[width=\textwidth,trim={2cm 13cm 18cm 18cm},clip]{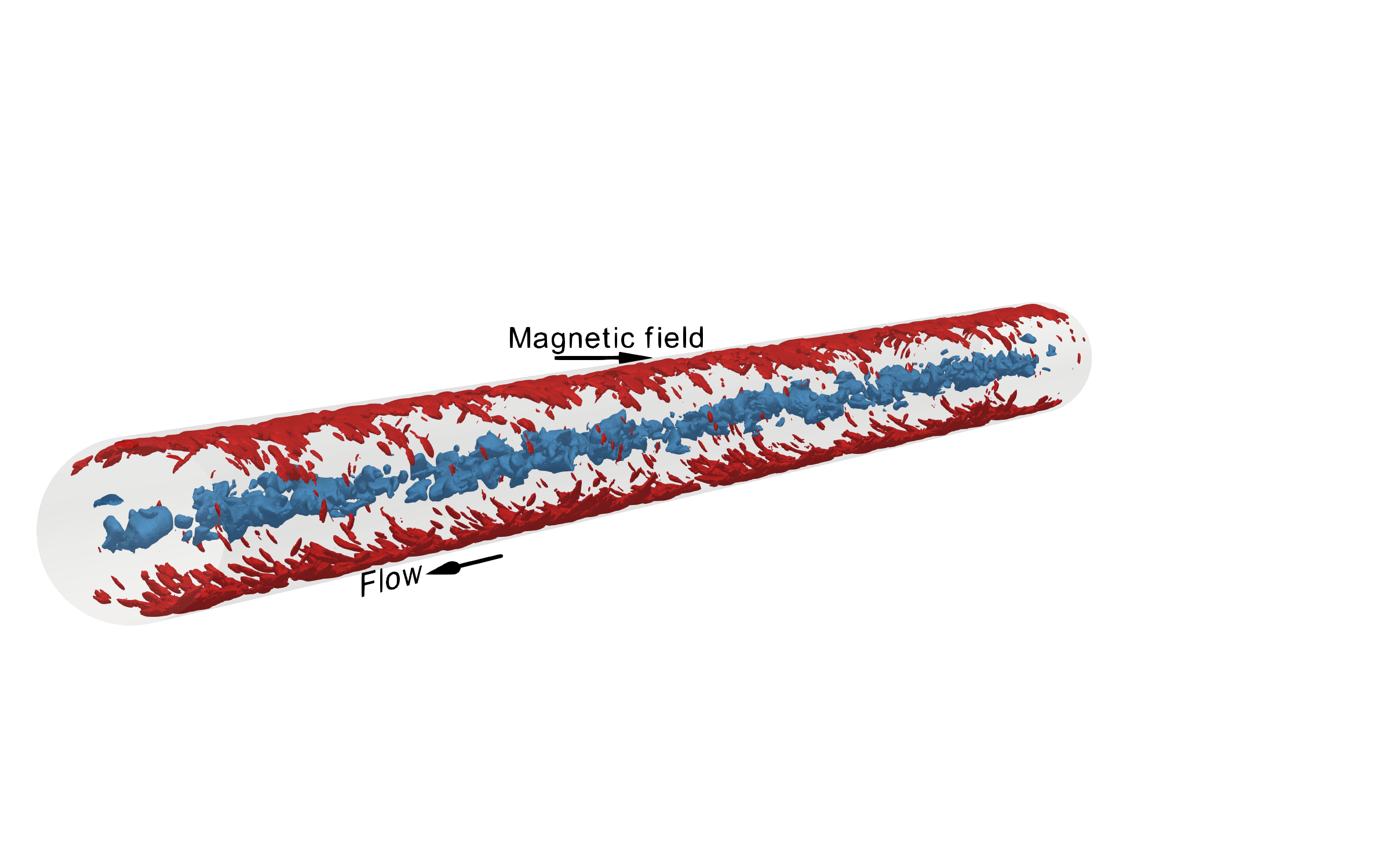}
    \put(-170,53){$t=3t_\textrm{opt}$}
  \end{subfigure}
  \caption{Several stages of nonlinear evolution of the streamwise component of the optimal perturbation for $Re=5000$, $Ha=5$, $\sigma_w=0$ and $e_0=10^{-3}$. The isosurfaces are shown for two wavelengths of the optimal perturbation.}
  \label{figure:nonlinear-evolved-contours-5-0-5000}
\end{figure}

In Fig.~\ref{figure:nonlinear-evolved-contours-5-0-5000}, we display different stages of the nonlinear evolution of the global optimal perturbation for $Ha=5$ and with $e_0 = 10^{-3}$. For small $t$, the nonlinear effects are negligible, and we observe the formation of streaks via the lift-up effect as predicted by the linear theory. At $t=t_\textrm{opt}$, nonlinear effects have significantly deformed the streaks. Within the center of the pipe, the flow structures form a series of arrow-like patterns tilted approximately by $45\%$ relative to each other. At later times, the trailing edges of the flow structures transition first before the flow becomes turbulent throughout its entire cross-section (note that only two isosurfaces of streamiwse velocity are shown).

\begin{figure}
  \centering
  \begin{subfigure}[t]{0.45\textwidth}
      \centering
      \includegraphics[width=\textwidth,trim={5cm 2cm 24cm 14cm},clip]{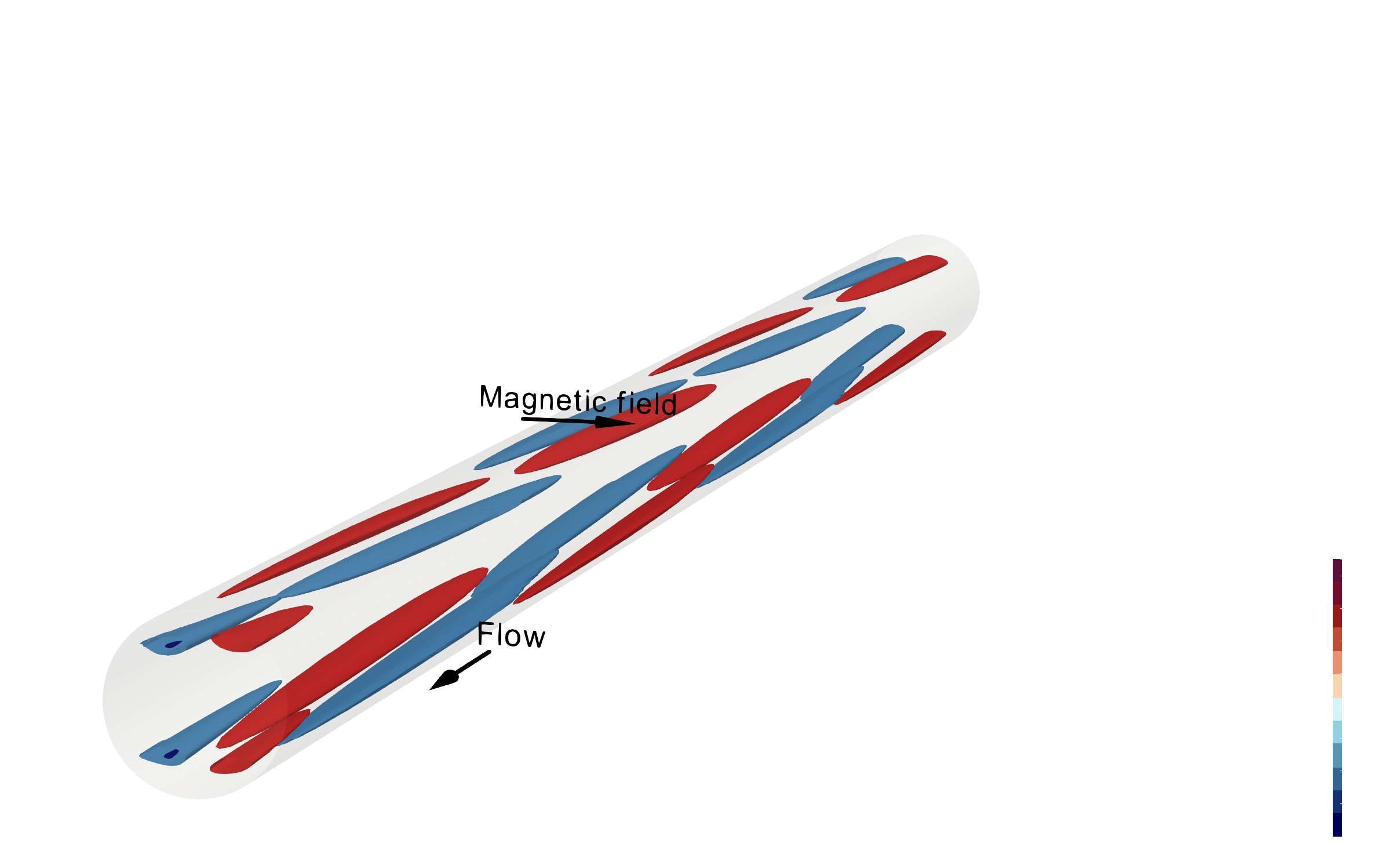}
      \put(-183,80){$t=0.1 t_\textrm{opt}$}
  \end{subfigure}
  \begin{subfigure}[t]{0.45\textwidth}
    \centering
    \includegraphics[width=\textwidth,trim={5cm 2cm 24cm 14cm},clip]{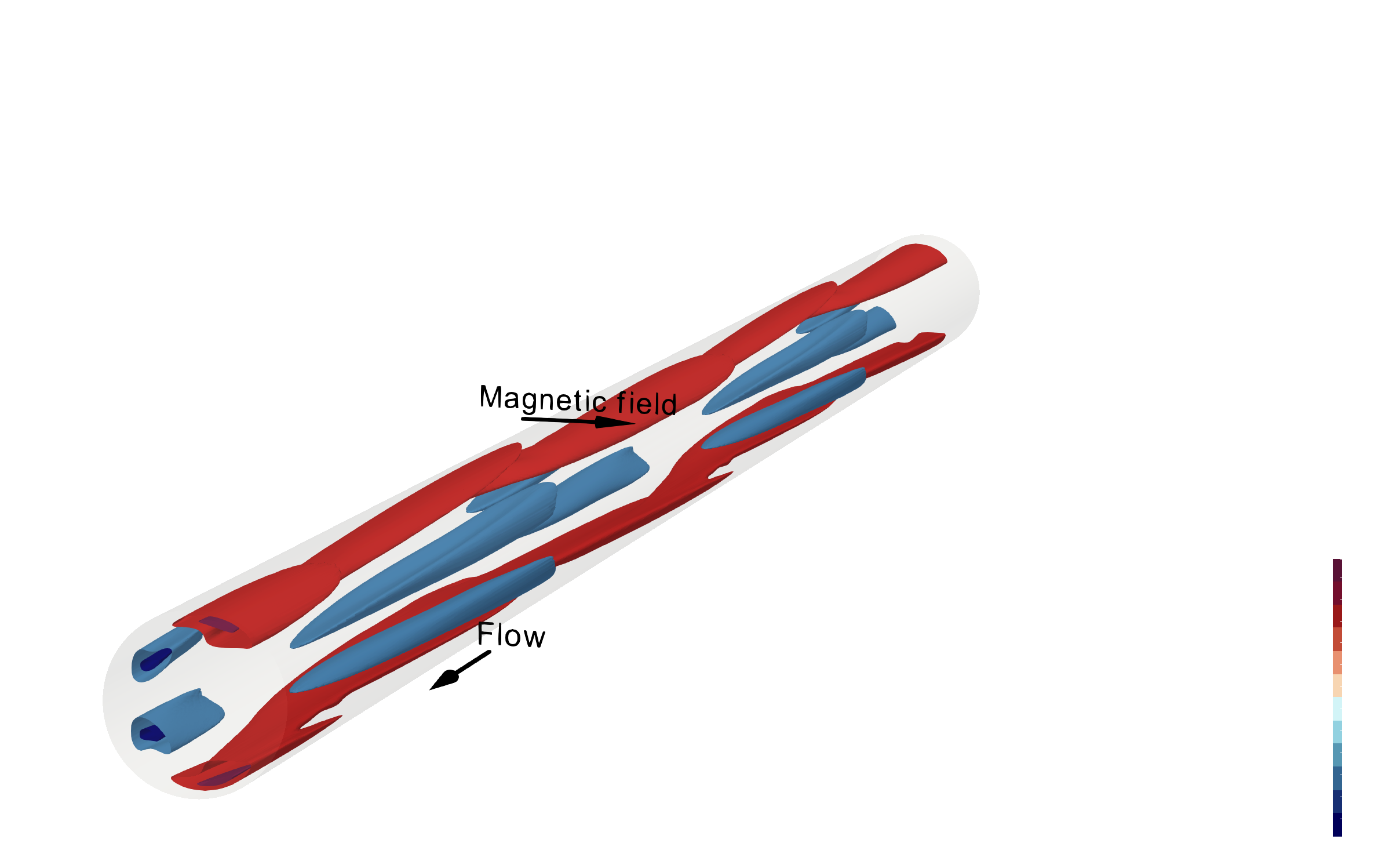}
    \put(-170,80){$t=t_\textrm{opt}$}
  \end{subfigure}
  \vskip\baselineskip
  \begin{subfigure}[t]{0.45\textwidth}
    \centering
    \includegraphics[width=\textwidth,trim={5cm 2cm 24cm 14cm},clip]{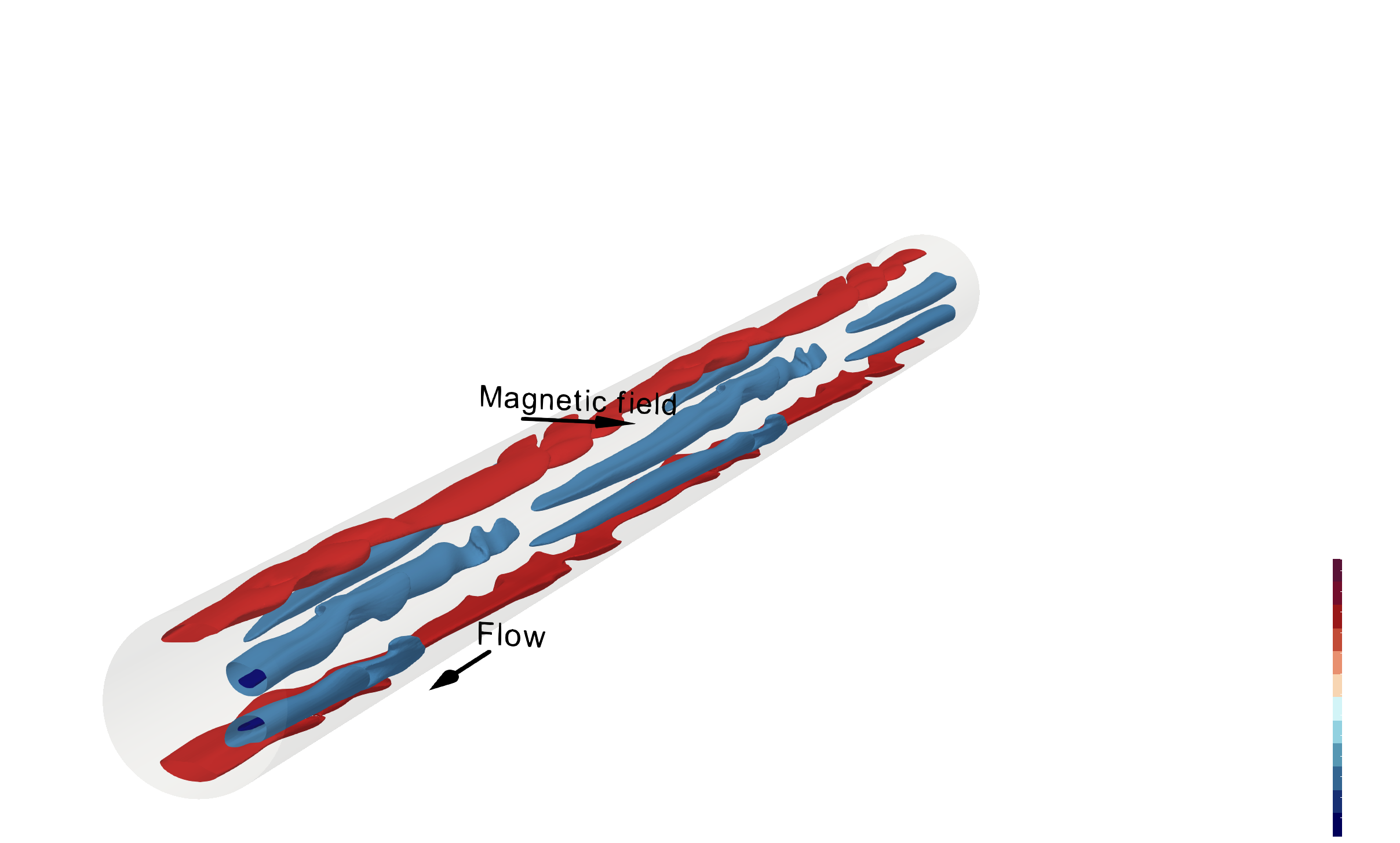}
    \put(-183,80){$t=1.35 t_\textrm{opt}$}
  \end{subfigure}
  \begin{subfigure}[t]{0.45\textwidth}
    \centering
    \includegraphics[width=\textwidth,trim={5cm 2cm 24cm 14cm},clip]{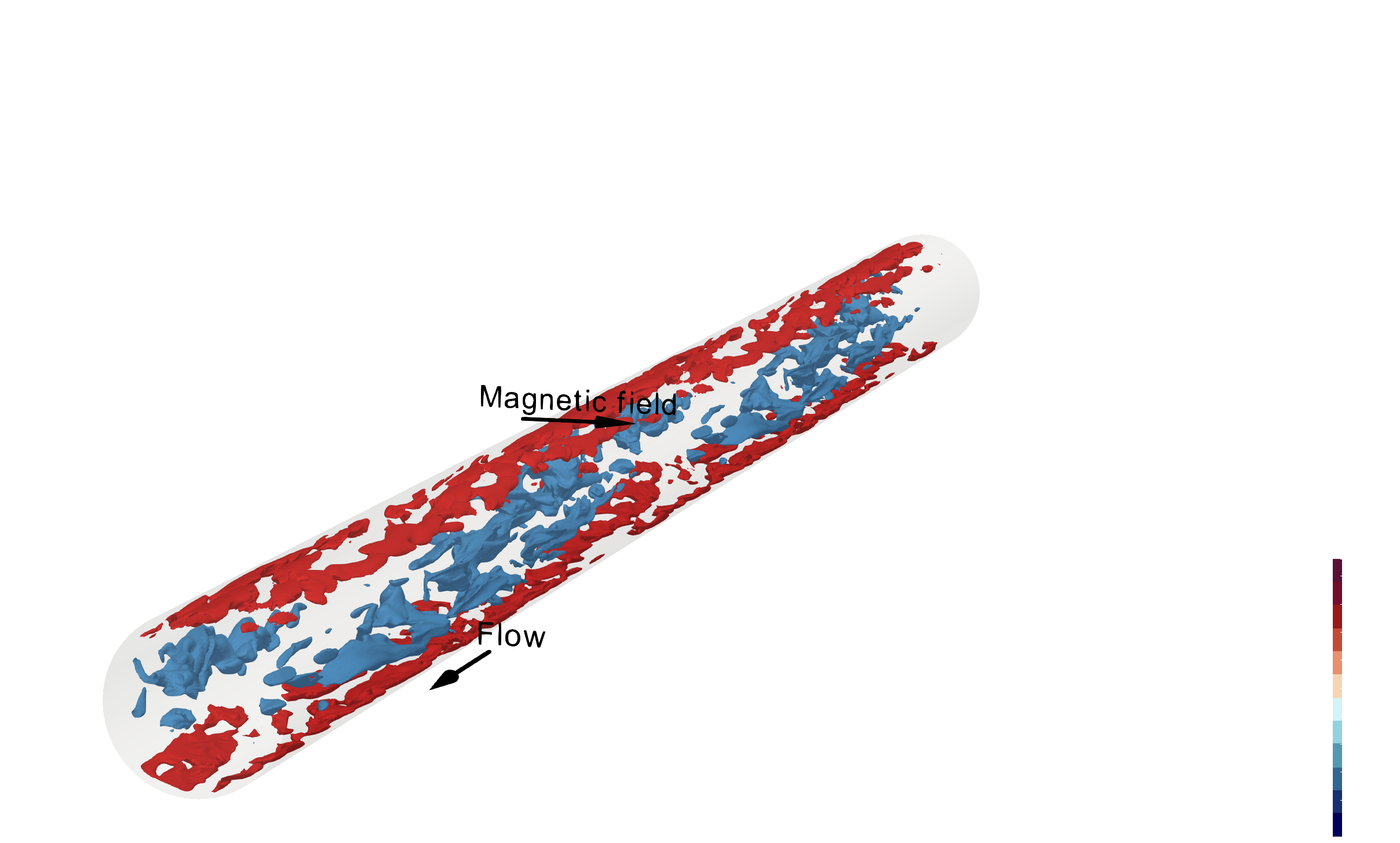}
    \put(-170,80){$t=2t_\textrm{opt}$}
  \end{subfigure}
  \caption{Stages of nonlinear evolution of the streamwise component of the optimal perturbation for $Re=5000$, $Ha=10$, $\sigma_w=0$ and $e_0=2 \cdot 10^{-4}$. The isosurfaces are shown for two wavelengths of the optimal perturbation.}
  \label{figure:nonlinear-evolved-contours-10-0-5000}
\end{figure}

The snapshots illustrating the stages of nonlinear evolution of the global optimal perturbation for $Ha=10$ and $e_0 = 2\cdot 10^{-4}$ are displayed in Fig.~\ref{figure:nonlinear-evolved-contours-10-0-5000}. At $t=0.1 t_\textrm{opt}$, we observe the initial formation of oblique streaks by means of the lift-up effect. At $t=t_\textrm{opt}$, they have been subjected to a secondary amplification and tilting downstream by the Orr-mechanism. Additionally, the streaks have already been modified by nonlinear interactions and the perturbation flow is no longer characterized by the two-plane symmetry, although the symmetry with respect to the direction of the magnetic field is preserved. Similar to the case of $Ha=5$, the destabilization of the streaks initially occurs at their trailing edge until eventually, the flow becomes fully turbulent.
\section{Conclusions}\label{sec:conclusions}
We have analyzed the transient growth of three-dimensional perturbations in the MHD pipe flow with transverse magnetic field. The main objective of this work was to understand the nature of optimal modes, identify the underlying mechanisms of their growth and provide some insight into their nonlinear evolution. In this study, we focused on flows at $Re=5000$, close to the transitional regime for low to moderate Hartmann numbers. Two types of electromagnetic boundary conditions at the pipe's wall were considered: perfectly insulating and perfectly conducting.

The results demonstrate that transient growth is progressively suppressed when increasing the intensity of the applied magnetic field. This behavior is clearly a consequence of Joule damping and is more pronounced in the case of a perfectly conducting pipe wall. This reduction in transient growth is accompanied by a gradual decrease of the optimal time at which the maximum amplification is reached and an increase of the initial growth rate of optimal perturbations. In contrast to pipe Poiseuille flow, the global maximum growth is achieved for disturbances with non-zero axial wavenumber, with this effect being more pronounced for a perfectly conducting pipe wall.

Depending on the Hartmann number, we have identified four types of optimal perturbations that are largely observed in similar ranges of Hartmann numbers in the perfectly insulating and perfectly conducting cases.

For weak magnetic fields ($Ha \lesssim 3$), the disturbance achieving the global maximum amplification is very similar to the hydrodynamic one, except for a slow modulation in the streamwise direction. This behavior has already been reported in Ref.~\cite{CassellsVoPotheratSheard2019} for the MHD duct flow.

In the range of Hartmann numbers $3 \lesssim Ha \lesssim 10$, the optimal perturbation
consists of a quartet of streamwise vortices occupying a large fraction of the pipe's cross-section. Over time, the perturbation of the streamwise velocity component gains significant kinetic energy via the lift-up effect and the initial streamwise vortices transform into high and low-velocity longitudinal streaks.

For moderate Hartmann numbers $10 \lesssim Ha \lesssim 50$ in the perfectly insulating case and $10 \lesssim Ha \lesssim 75$ in the perfectly conducting case, the optimal perturbation emerges in the form of oblique vortices, localized in the Roberts layers of the flow. They appear to be elongated in the direction of the magnetic field and tilted against the direction of the base flow. As we increase the Hartmann number, their tilting becomes more pronounced. The lift-up effect remains the primary mechanism of growth for this type of disturbance, leading to the occurrence of oblique streaks. However, these perturbations are further amplified in a second stage by the Orr-mechanism, resulting in the final tilting of the perturbations in the same direction as the mean flow. This second-stage growth affects all velocity components and becomes proportionally more significant with increasing Hartmann numbers, as the lift-up effect loses importance.

Lastly, for $Ha\gtrsim 75$ in the perfectly insulating case and $Ha\gtrsim 100$ in the perfectly conducting case, the optimal disturbance takes form of spanwise vortices, localized in the Roberts layers and tilted against the base flow. The emergence of such perturbations is consistent with the action of a strong magnetic field, which suppresses velocity variations along its direction and promotes quasi-two-dimensional motion. Moreover, an optimal perturbation of similar structure has been described in Ref.~\cite{CassellsVoPotheratSheard2019} in the case of the MHD duct flow with an intense transverse magnetic field. In this regime, the growth of the disturbance is achieved solely by means of the Orr-mechanism. Over time, the streamlines contained in the opposite Roberts layers merge, resulting in the creation of large-scale vortices that occupy the entire cross-section of the pipe. Qualitatively, the optimal perturbations in this regime evolve in a manner almost identical to that described in Ref.~\cite{Farrell1988} for purely 2D plane Poiseuille flow.

We have also addressed the nonlinear evolution of the optimal perturbations. For $Ha = 5$, transition to turbulence has been observed for a fraction of initial kinetic energy of disturbance $e_0$ larger than $10^{-3}$ (without the addition of 3D random noise), whereas for $Ha = 10$, sustained turbulence could be initiated for $e_0 > 10^{-4}$. In both cases, the transition is achieved by a secondary instability of the streaks generated via the lift-up effect. In the case $Ha=10$, however, the lift-up effect is complemented by a second-stage amplification by the Orr-mechanism before the transition is observed. At $Ha=5$, we have also considered the nonlinear evolution of the (local) optimal perturbation at $t=55$. Although it achieves only $88\%$ of the global maximum amplification, it allows a more efficient path to turbulence through its intrinsic three-dimensional nature. This stress the fact that although global optimal perturbations achieve the global maximum amplification through transient growth, some less amplified perturbations with larger initial growth rate may provide better pathways to turbulence.

For $Ha \geq 20$, no turbulence could be observed, even with the addition of three-dimensional random noise to the modulated flow. We note that in previous DNS of the turbulent MHD pipe flow with transverse magnetic field at $Re=5000$, full relaminarization occurred for $Ha \ge 23$ \cite{KrasnovThessBoeckZhaoZikanov2013}. This discrepancy may originate from several factors. First, although the parameters of transition to turbulence and relaminarization of an initially turbulent flow typically lie in a similar range, they often do not coincide. More importantly, in Ref.~\cite{KrasnovThessBoeckZhaoZikanov2013}, the authors observed that for $18 \le Ha \le 23$, turbulence did not occupy the entire pipe but occurred in patches, surrounded by a laminar flow. The realization of patterned turbulence typically requires particularly large computational domains and long simulation times, which have not been achieved in this study. To achieve a turbulent state for Hartmann numbers beyond that range, higher Reynolds numbers are necessary, which, in turn, demand significantly more computational resources. In particular, we could not explore the nonlinear dynamics of the quasi-2D regime observed in the transient growth analysis for $Ha \gtrsim 75$ in the perfectly insulating case. It is nevertheless a likely important regime in high-Hartmann number applications like fusion energy, and worthy of future investigations.

\begin{acknowledgments}
  The authors are grateful to A. Potherat for fruitful discussions.
  Computational resources have been provided by the Consortium des Équipements de Calcul Intensif (CÉCI), funded by the Fonds de la Recherche Scientifique de Belgique (F.R.S.-FNRS) under Grant No. 2.5020.11 and by the Walloon Region.
  This work has been carried out within the framework of 1) the EUROfusion Consortium, funded by the European Union via the Euratom Research and Training Programme (Grant Agreement No 101052200 — EUROfusion), 2) the Belgian Fusion Association and has received funding from the FPS Economy, SMEs, Self-Employed and Energy. Views and opinions expressed are however those of the author(s) only and do not necessarily reflect those of the European Union or the European Commission nor of the FPS Economy, SMEs, Self-Employed and Energy. Neither the European Union, the European Commission nor the FPS Economy, SMEs, Self-Employed and Energy can be held responsible for them.
  \end{acknowledgments}

\bibliography{ref}

\end{document}